\documentclass[pra,aps,showpacs,onecolumn,twoside,superscriptaddress]{revtex4}

\usepackage{amsmath,amsfonts,amssymb,color,epsfig,graphics,graphicx,latexsym,revsymb,theorem,verbatim}

\newtheorem{definition}{Definition}
\newtheorem{proposition}[definition]{Proposition}
\newtheorem{lemma}[definition]{Lemma}

\newtheorem{theorem}[definition]{Theorem}
\newtheorem{corollary}[definition]{Corollary}
\newtheorem{conjecture}[definition]{Conjecture}

\newtheorem{remark}[definition]{Remark}

\def\squareforqed{\hbox{\rlap{$\sqcap$}$\sqcup$}}
\def\qed{\ifmmode\squareforqed\else{\unskip\nobreak\hfil
\penalty50\hskip1em\null\nobreak\hfil\squareforqed
\parfillskip=0pt\finalhyphendemerits=0\endgraf}\fi}
\def\endenv{\ifmmode\;\else{\unskip\nobreak\hfil
\penalty50\hskip1em\null\nobreak\hfil\;
\parfillskip=0pt\finalhyphendemerits=0\endgraf}\fi}
\newenvironment{proof}{\noindent \textbf{{Proof.~} }}{\qed}

\def\bcj{\begin{conjecture}}
\def\ecj{\end{conjecture}}
\def\bcr{\begin{corollary}}
\def\ecr{\end{corollary}}
\def\bd{\begin{definition}}
\def\ed{\end{definition}}
\def\bea{\begin{eqnarray}}
\def\eea{\end{eqnarray}}
\def\bem{\begin{enumerate}}
\def\eem{\end{enumerate}}
\def\bim{\begin{itemize}}
\def\eim{\end{itemize}}
\def\bl{\begin{lemma}}
\def\el{\end{lemma}}
\def\bpf{\begin{proof}}
\def\epf{\end{proof}}
\def\bpp{\begin{proposition}}
\def\epp{\end{proposition}}
\def\br{\begin{remark}}
\def\er{\end{remark}}
\def\bt{\begin{theorem}}
\def\et{\end{theorem}}

\def\a{\alpha}
\def\b{\beta}
\def\g{\gamma}
\def\d{\delta}
\def\e{\epsilon}

\def\z{\zeta}

\def\t{\theta}
\def\i{\iota}

\def\l{\lambda}
\def\m{\mu}
\def\n{\nu}
\def\x{\xi}

\def\r{\rho}
\def\s{\sigma}

\def\ph{\varphi}
\def\c{\chi}
\def\ps{\psi}

\def\G{\Gamma}
\def\D{\Delta}

\def\S{\Sigma}

\def\GL{{\mbox{\rm GL}}}
\def\PGL{{\mbox{\rm PGL}}}

\def\Un{{\mbox{\rm U}}}

\newcommand{\nc}{\newcommand}

\nc{\cA}{{\cal A}} \nc{\cB}{{\cal B}} \nc{\cC}{{\cal C}}
\nc{\cD}{{\cal D}} \nc{\cE}{{\cal E}} \nc{\cF}{{\cal F}}
\nc{\cG}{{\cal G}} \nc{\cH}{{\cal H}} \nc{\cI}{{\cal I}}
\nc{\cJ}{{\cal J}} \nc{\cK}{{\cal K}} \nc{\cL}{{\cal L}}
\nc{\cM}{{\cal M}} \nc{\cN}{{\cal N}} \nc{\cO}{{\cal O}}
\nc{\cP}{{\cal P}} \nc{\cR}{{\cal R}} \nc{\cS}{{\cal S}}
\nc{\cT}{{\cal T}} \nc{\cU}{{\cal U}} \nc{\cX}{{\cal X}}
\nc{\cZ}{{\cal Z}}

\def\diag{\mathop{\rm diag}}
\def\dim{\mathop{\rm Dim}}

\def\lin{\mathop{\rm span}}

\def\ppt{\mathop{\rm PPT}}

\def\rank{\mathop{\rm rank}}

\def\sep{\mathop{\rm SEP}}

\def\tr{\mathop{\rm Tr}}

\def\bigox{\bigotimes}
\def\dg{\dagger}

\def\ox{\otimes}

\newcommand{\bra}[1]{\langle#1|}
\newcommand{\ket}[1]{|#1\rangle}
\newcommand{\proj}[1]{| #1\rangle\!\langle #1 |}
\newcommand{\ketbra}[2]{|#1\rangle\!\langle#2|}
\newcommand{\braket}[2]{\langle#1|#2\rangle}

\newcommand{\jmp}{J. Math. Phys.}
\newcommand{\jpa}{J. Phys. A}

\newcommand{\pla}{Phys. Lett. A}

\font\germ=eufm10

\def\gm{{\mbox{\germ m}}}

\def\bC{{\mbox{\bf C}}}

\begin{document}
\title{Description of rank four PPT entangled states of two qutrits}

\author{Lin Chen}
\affiliation{Centre for Quantum Technologies, National University of
Singapore, 3 Science Drive 2, Singapore 117542}
\email{cqtcl@nus.edu.sg (Corresponding~Author)}

\def\Dbar{\leavevmode\lower.6ex\hbox to 0pt
{\hskip-.23ex\accent"16\hss}D}
\author {{ Dragomir {\v{Z} \Dbar}okovi{\'c}}}

\affiliation{Department of Pure Mathematics and Institute for
Quantum Computing, University of Waterloo, Waterloo, Ontario, N2L
3G1, Canada}
\email{djokovic@uwaterloo.ca}

\begin{abstract}
It is known that some two qutrit entangled states of rank 4 with
positive partial transpose (PPT) can be built from the unextendible
product bases (UPB) [ \prl {\bf82}, 5385 (1999) ]. We show that this
fact is indeed universal, namely all such states can be constructed
from UPB as conjectured recently by Leinaas, Myrheim and Sollid.
We also classify the 5-dimensional subspaces of two qutrits which
contain only finitely many product states (up to scalar multiple),
and in particular those spanned by a UPB.
\end{abstract}

\date{ \today }




\maketitle

\tableofcontents


\section{Introduction}

The positive-partial-transpose entangled states (PPTES) are of
particular importance and interest in quantum information (for a
review see \cite{hhh09}). For a state $\r$ acting on the Hilbert
space $\cH:=\cH_A \ox \cH_B$, the partial transpose computed in an orthonormal (o.n.) basis $\{\ket{a_i}\}$ of system A, is defined by
$\r^\G :=\sum_{ij} \bra{a_i} \r \ket{a_j} \ketbra{a_j}{a_i}$.
We say that $\r$ is a PPT [NPT] state if $\r^\G\ge0$ [$\r^\G \ngeq 0$,
i.e., $\r^\G$ has at least one negative eigenvalue].
The most intriguing feature of PPTES is their non-distillability
under local operations and classical communications (LOCC)
\cite{hhh98}. This means that a PPTES, say $\r$, cannot be
locally converted (asymptotically) into a pure entangled state
even if infinitely many copies of $\r$ are provided.
Since most quantum-information tasks require pure entangled states,
a PPTES is a quantum resource which cannot be directly used in
practice \cite{bds96}. Nevertheless in the past few years the PPTES
have been extensively studied in connection with the phenomena
of entanglement activation and universal usefulness
\cite{hhh99,masanes06}, the distillable key \cite{hhh05}, the
symmetry permutations \cite{tg09} and entanglement witnesses
\cite{lkc00}, both in theory and experiment.

For many applications, it is an important and basic problem to
decide whether a given PPT state is entangled or separable, i.e.,
the convex sum of product states $\sum_i \proj{a_i,b_i}$
\cite{werner89}. The separability problem has an extensive
application in quantum information, metrology, computing, quantum
non-locality, and mathematics (like positive maps and
$C^*$-algebras). Moreover, the problem has been proved to be NP-hard
and hence it attracted a lot of attention from computer scientists
\cite{gurvits03}. In 1996, the first necessary condition was given
by Peres \cite{peres96}, saying that the separable states are always
PPT. So to solve the problem it suffices to consider only the PPT
states. Next in 1997, the Horodeckis \cite{hhh96} showed that this
is necessary and sufficient for $2\ox 2$ and $2\ox 3$ states.
However both of these cases lack the PPTES. Actually since the first
PPTES was constructed \cite{horodecki97}, researchers lacked for a
long time the analytical characterization of PPTES in any bipartite
systems of given rank and local dimensions. For example, it was
surprisingly difficult to decide whether a given state of rank 4 in
$3\ox 3$ space is a PPTES, which is also the smallest space in which
PPTES may exist \cite{hst99}. A well-known method \cite{bdm99} for
construction of PPTES proposed in 1999 was based on the unextendible
product bases (UPB). It is applicable to arbitrary bipartite and
multipartite quantum systems. Another systematic method for
two-qutrit systems was provided by Chen and {\Dbar}okovi{\'c} who
proved in 2010 that a PPT state of rank 4 is entangled if and only
if there is no product state in its range \cite{cd11}.

Our main result (see Theorem \ref{thm:maintheorem=PPT3x3rank4})
shows that any two-qutrit PPTES $\r$ of rank 4 can be
constructed from an unextendible product basis (UPB) \cite{bdm99}
by using the method proposed by Leinaas, Myrheim and
Sollid in \cite{slm11}. We state their conjecture formally as
Conjecture \ref{conj:HipLMS}.
Explicitly, we prove that (up to normalization)
$\r=A\ox B~\Pi\{\psi\}~ A^\dag\ox B^\dag$,
where $\{\psi\}$ is a UPB, $\Pi\{\psi\}$
is the projector on the subspace orthogonal to $\{\psi\}$, and
$A,B\in\GL_3$, the group of invertible complex matrices of order 3.
Let us point out that the papers \cite{hhm11,slm11} provided
strong numerical evidence for the validity of this result and
motivated us to pursue this study.
Moreover, the authors of these papers suggest that the higher
dimensional PPTES may have similar properties,
although some of them have to be modified.
This may be of interest for further research in this direction.

It is well-known that the set $\cS_{\sep}$ of separable states and
the set $\cS_{\ppt}$ of PPT states are both compact and convex, and
that $\cS_{\sep} \subseteq \cS_{\ppt}$. A basic task is then to
characterize their extreme points. The latter are the states which
are not convex sums of other states in the convex set.
It is also a well-known fact that we can generate any state in such a set by taking the convex sum of extreme points.
Though it is known that the extreme points of $\cS_{\sep}$ are
exactly the pure product states \cite{werner89}, we know quite little
about the extreme points of $\cS_{\ppt}$ \cite{lmo07}.
Only in the case of $2\ox 4$ systems,
the PPTES have been partially classified by the ranks of both
the state and its partial transpose \cite{agk10}. Here, we will show
that all two-qutrit PPTES of rank 4 are such extreme points
(and also edge PPTES).

We also show that no PPTES of rank 4 exist in the symmetric
subspace of the two-qutrit system.
Furthermore, it is known that the UPB basis states are not
distinguishable under LOCC \cite{bdf99}. Thus we exhibit the
essential connection between the LOCC-indistinguishability and PPTES.

The content of our paper is as follows. In Sec. \ref{sec:ProdV}
we define the notion of general position for $m$-tuples of
product vectors (or points on the Segre variety $\cP^2\times\cP^2$),
see Definitions \ref{def:GenPosition} and \ref{def:SigmaGenPos}.
We also define and study the properties of the biprojective
(BP) equivalence of such $m$-tuples, and in particular for
quintuples. To a quintuple of product vectors in general position
we assign six
$\GL_3\times\GL_3$ invariants $(J_i^A,J_i^B)$, $(i=1,2,3)$,
and show that two quintuples are BP-equivalent if and only if
they share the same invariants.
These invariants are subject to the relations
$J_1^AJ_2^AJ_3^A=1$ and $J_1^BJ_2^BJ_3^B=1$, and can take any
complex values except 0 and 1. We then investigate the product
states contained in the 5-dimensional subspace spanned by a
quintuple of product states in general position.
In particular, it follows form the part(a) of
Proposition \ref{prop:5generalposition} that
if $\r$ is a two-qutrit PPT state of rank 4 then its kernel
contains only finitely many product vectors (up to scalar multiple).
This fact is used later to prove
Theorem \ref{thm:3x3rank4PPTES,6productstates}.

In Sec. \ref{sec:prodstate=3x3dim5} we consider the 8-dimensional
complex projective space associated to the $3\ox 3$ Hilbert space
$\cH$.
We consider the case where a 4-dimensional projective subspace,
$\cP^4$, and the Segre variety $\S_{2,2}$ intersect only at
finitely many points.
Then there are at most 6 points of intersection and to each of
them one assigns (in Algebraic Geometry) a positive integer
known as the intersection multiplicity.
The sum of these integers is necessarily equal to 6.
We refer to the corresponding partition of 6 as the
intersection pattern of $\cP^4$ and $\S_{2,2}$.
We show that all feasible patterns, i.e., all partitions of 6, occur. For instance, there exists a 5-dimensional vector subspace of
$\cH$ which contains only one product state (up to a scalar multiple).

In Sec. \ref{sec:equivalentUPB} we study the invariants of
quintuples of product states in general position which arise from UPB.
These invariants are real and take values in one of the open
intervals $N:=(-\infty,0)$, $p:=(0,1)$ and $P:=(1,+\infty)$.
By replacing the values of the 6 invariants by the letter
designating the interval to which the invariant belongs, we obtain
the 6-letter symbol attached to the quintuple. We show that
among the 120 permutations of the 5 product states of a UPB,
there are exactly 12 different symbols that arise in this way.
We say that these 12 symbols are UPB symbols.
The same assertion is valid if we use all 6 product states
contained in the 5-dimensional subspace spanned by a UPB and
construct from them all 720 possible quintuples. Their symbols
are also UPB symbols. This is the key tool in the proof of
our main result.

In Sec. \ref{sec:product=3x3rank4PPT} we prove that the kernel of
any PPTES $\r$ of rank 4 contains exactly 6 product states
(up to scalar multiple) and that these 6 states are in general
position. Next we show that there exist $A,B\in\GL_3$ such that
the transformed state $\s:=A\ox B~\r~A^\dag\ox B^\dag$ is
invariant under partial transpose, i.e., $\s^\G=\s$.
We also show that if two normalized PPTES of rank 4 have
the same range (or kernel) then they are equal.
Finally, by making use of the UPB symbols we
prove our main result, Theorem \ref{thm:maintheorem=PPT3x3rank4}
We conclude that section with a description of the stabilizer
$G_\r$ of a PPTES $\r$ of rank 4 in the group
$\PGL:=\PGL_3\times\PGL_3$. We show that this
is a finite group isomorphic to a subgroup of the symmetric group
$S_6$ of degree 6, and we compute these groups for the two most
interesting and most symmetric cases arising from {\bf Pyramid}
and {\bf Tiles} UPB.

In Sec. \ref{sec:PhysicalApplication} we discuss a few physical applications of our results. We show how to build analytically
the PPTES of rank 4 of two qutrits. This is also helpful for the
construction of entanglement witnesses which detect these PPTES.

Throughout the paper we use the expression ``$M\times N$ state''
to mean a bipartite state, say $\r$, whose local ranks are
$\rank \r_A= M$ and $\rank\r_B=N$. By $\cR(\r)$ we denote the 
range of a linear operator $\r$.

Another proof of the Conjecture \ref{conj:HipLMS} was obtained by
L. Skowronek \cite{skowronek11}.

\section{Product states of two qutrits} \label{sec:ProdV}

For convenience, we shall represent any product vector $\ket{x,y}$
as a $3\times3$ matrix $[x_iy_j]$ where $i,j\in\{1,2,3\}$. In this
notation, product vectors correspond to matrices of rank 1. We shall
often work with these vectors only up to scalar multiple, in which
case we consider them as points in the 8-dimensional complex
projective space $\cP^8$ associated to the 9-dimensional Hilbert
space $\cH=\cH_A\ox\cH_B$. Then the product vectors form the so
called Segre variety $\S_{2,2}=\cP^2\times\cP^2\subseteq\cP^8$,
isomorphic to the product of two copies of the complex projective
plane $\cP^2$.

\subsection{Projective invariants $J_1,J_2,J_3$}
The complex general linear group in dimension 3, that is the group
of invertible complex matrices of order 3, will be denoted
by $\GL_3$. We shall also use the group $\GL:=\GL_3\times\GL_3$,
the direct product of two copies of $\GL_3$.
It acts naturally on $\cH=\cH_A\ox\cH_B$ via invertible linear
operators (ILO) $A\ox B$.
The points of the complex projective plane $\cP^2$ will be
identified with the 1-dimensional subspaces of $\cH_A$ (or $\cH_B$).
As $\GL_3$ permutes the 1-dimensional subspaces of $\cH_A$,
it induces an action on $\cP^2$. The subgroup $\bC^\times$,
consisting of nonzero scalar matrices of $\GL_3$, acts trivially
on $\cP^2$. Thus we obtain an action of the quotient group
$\PGL_3:=\GL_3/\bC^\times$ on $\cP^2$. This group is known as
the complex projective general linear group (in dimension 3)
and the transformations that it induces on $\cP^2$ are known as
projective transformations.
The direct product $\PGL:=\PGL_3\times\PGL_3$ acts on the
Segre variety $\cP^2\times\cP^2$, and we refer to it as the
group of biprojective transformations of $\S_{2,2}$.

Invertible linear operators are also useful in quantum information.
Two $n$-partite states $\r$ and $\s$ are {\em equivalent under stochastic LOCC} (or {\em SLOCC-equivalent}) if
$\r=\bigox^n_{i=1} A_i \s \bigox^n_{i=1}A_i^\dg$ for some ILO
$A_1,\ldots,A_n$. They are {\em LU-equivalent} if $A_i$ can be
chosen to be unitary. In most cases of the present work,
we will have $n=2$. Both LOCC and SLOCC are
referred to as physical operations in quantum information
\cite{hhh09}. The essential difference between them is that LOCC
can be implemented with certainty while SLOCC succeeds only with
some nonzero probability.

 \bd
\label{def:GenPosition} We say that an $m$-tuple of pure states
$(\ket{\phi_k})_{k=0}^{m-1}$ in $\cH_A$ (or $\cH_B$) is in {\em
general position} if any two or three of them are linearly
independent. If $(\ket{\phi_k})_{k=0}^{m-1}$ and
$(\ket{\phi'_k})_{k=0}^{m-1}$ are two $m$-tuples in $\cH_A$ such
that $A\ket{\phi_k}\propto\ket{\phi'_k}$ for some invertible matrix
$A$, then we say that they are {\em projectively equivalent} or {\em
P-equivalent}. \ed

Let us recall the following elementary fact from Linear Algebra also
known as the Four Point Lemma (see e.g. \cite[Lemma 11.2]{Gib}): \bl
\label{le:FPL} If $(\ket{\phi_k})_{k=0}^3$ and
$(\ket{\phi'_k})_{k=0}^3$ are quadruples of pure states in general
position in $\cH_A$, then there exists an invertible matrix $A$,
unique up to a scalar factor, such that
$A\ket{\phi_k}\propto\ket{\phi'_k}$ for each $k$. \el

In particular, for any quadruple $(\ket{\phi_k})_{k=0}^3$ in general
position there exists an invertible matrix $A$ such that
$A\ket{\phi_k}\propto\ket{k}$ for $k=0,1,2$ and
$A\ket{\phi_3}=\ket{0}+\ket{1}+\ket{2}$. For convenience, we say
that $A$ transforms the quadruple $(\ket{\phi_k})_{k=0}^3$ into the
{\em canonical form}. One can construct $A$ explicitly as follows.
Let $X=[\ket{\phi_0}~\ket{\phi_1}~\ket{\phi_2}]$ and let $D$ be the
diagonal matrix whose diagonal entries are the components of the
vector $X^{-1}\ket{\phi_3}$. Then we have $A=D^{-1}X^{-1}$.

In the language of Projective Geometry the above lemma can be
restated as follows. If $(P_k)_{k=0}^3$ and $(P'_k)_{k=0}^3$ are two
quadruples of points in the complex projective plane $\cP^2$ in
general position (i.e., they are distinct and no three points $P_k$
are colinear, and similarly for the points $P'_k$) then there is
a projective transformation $T$ of $\cP^2$ such that $T(P_k)=P'_k$
for each $k$.

However, the action of $\GL_3$ on quintuples of points in general
position in $\cP^2$ is not transitive. Indeed let
$(P_k=\ket{\phi_k})_{k=0}^4$ be such a quintuple. Then all
determinants
 \bea
\D_{i,j,k}=\det[\ket{\phi_i}\ \ket{\phi_j}\ \ket{\phi_k}], \quad
(i,j,k \text{ distinct),}
 \eea
are nonzero. The rational functions \bea \label{J1-J2} J_1 =
\frac{\D_{2,0,4}\D_{0,1,3}}{\D_{2,0,3}\D_{0,1,4}},\quad J_2 =
\frac{\D_{0,1,4}\D_{1,2,3}}{\D_{0,1,3}\D_{1,2,4}},\quad J_3 =
\frac{\D_{1,2,4}\D_{2,0,3}}{\D_{1,2,3}\D_{2,0,4}} \eea are
projective $\GL_3$-invariants of quintuples in general position.
These invariants may take arbitrary complex values, except 0 and 1,
subject to the relation $J_1J_2J_3=1$. The following result follows
easily from the Four Point Lemma and the fact that $P_4$ is
determined uniquely by the quadruple $(P_k)_{k=0}^3$ and the values
of the invariants $J_i$.

 \bpp
\label{prop:5-Points} Two quintuples of points
$(P_k=\ket{\phi_k})_{k=0}^4$ and $(P'_k=\ket{\phi'_k})_{k=0}^4$ in
general position are $P$-equivalent if and only if they share the
same values of the three invariants $J_i$. \epp

This means that if the two quintuples satisfy the invariance
conditions, then there exists $A\in\GL_3$ such that
$A\ket{\phi_k}=c_k\ket{\phi'_k}$ for some scalars $c_k$, which may
be all different.

\bd \label{def:SigmaGenPos}
We say that an $m$-tuple
$(\ket{\psi_k}=\ket{\phi_k}\ox\ket{\chi_k})_{k=0}^{m-1}$ of
non-normalized product states in a $3\times3$ system is in {\em
general position} if each of the $m$-tuples
$(\ket{\phi_k})_{k=0}^{m-1}$ and $(\ket{\chi_k})_{k=0}^{m-1}$ is in
general position. In that case we also say that the corresponding
$m$-tuple of points $(P_k=\ket{\psi_k})_{k=0}^{m-1}$ on $\S_{2,2}$
is in {\em general position}.
We also say that two $m$-tuples of product states
$(\ket{\psi_k}=\ket{\phi_k}\ox\ket{\chi_k})_{k=0}^{m-1}$ and
$(\ket{\psi'_k}=\ket{\phi'_k}\ox\ket{\chi'_k})_{k=0}^{m-1}$ are {\em
biprojectively equivalent} or {\em BP-equivalent} if there exists
$A\ox B\in\GL$ such that $(A\ox B)\ket{\psi_k}\propto\ket{\psi'_k}$
for each $k$.
\ed
We shall use the same terminology for the $m$-tuples of points
lying on $\S_{2,2}$.

In the important case $m=5$, we have
two sets of invariants, one for $(\ket{\phi_k})_{k=0}^4$ and the
other for $(\ket{\chi_k})_{k=0}^4$. We shall denote the former by
$J_i^A$ and the latter by $J_i^B$.

The following proposition is an immediate consequence of the
definitions and results stated above.
 \bpp \label{prop:Trans-4-5}
The group $\GL$ acts transitively on the quadruples of points in
general position in $\S_{2,2}$. Two quintuples of points
$(P_k=\ket{\phi_k}\ox\ket{\chi_k})_{k=0}^4$ and
$(P'_k=\ket{\phi'_k}\ox\ket{\chi'_k})_{k=0}^4$ in $\S_{2,2}$, both
in general position, are $BP$-equivalent if and only if the
quintuples of points $(\ket{\phi_k})_{k=0}^4$ and
$(\ket{\phi'_k})_{k=0}^4$ in $\cP^2$ are $P$-equivalent and the same
is true for the quintuples of points $(\ket{\chi_k})_{k=0}^4$ and
$(\ket{\chi'_k})_{k=0}^4$.
 \epp

We apply our results to the manipulation of families of quantum
states. That is, we consider the condition on which two sets of
quantum states $\{\r_1,\ldots,\r_n\}$ and $\{\s_1,\ldots,\s_n\}$ can
be (probabilistically) simultaneously convertible via a physical
operation $\e$, i.e., $\e(\r_i):=\sum_j A_j (\r_i) A_j^\dg=\s_i$,
for all $i$ \cite{cjw03}. The problem is in general difficult even
for the case of single-party and $n=2$, which has been studied in
terms of single-party states for distinguishing quantum operations
\cite{dfy09}. In present work, $\r_i$ and $\s_i$ are single-party
pure states in general position. To realize the conversion, the
Kraus operators $A_i$ have to be pairwise proportional. So the Four
Point Lemma and Proposition \ref{prop:5-Points} can be used to
decide the simultaneous conversion between two sets of 4 and 5
qutrit states in general position, respectively. Moreover,
Proposition \ref{prop:Trans-4-5} works for the conversion between
two sets of bipartite product states. Hence we have produced some
new manipulatable families of states. Further, we have

\bl \label{le:generic} If $(\ket{\psi_k})_{k=0}^3$ is a quadruple of
product states in general position, then the subspace that they span
contains no other product state. \el \bpf Let
$\ket{\psi_k}=\ket{\phi_k}\ox\ket{\chi_k}$, $k=0,\ldots,3$. By
applying the Four Point Lemma in $\cH_A$ and $\cH_B$, we may assume
that the quadruples $(\ket{\psi_k})_{k=0}^3$ are in the canonical
form
 \bea
\ket{\psi_k}\propto\ket{kk},\ (k=0,1,2), \quad \ket{\psi_3}\propto
\left[
\begin{array}{ccc}
1 & 1 & 1 \\ 1 & 1 & 1 \\ 1 & 1 & 1 \\
\end{array} \right].
 \eea
Note that the space of $3\times3$ diagonal matrices contains only 3
product vectors (up to a scalar multiple). Hence the assertion of
the lemma follows from the fact that a non-diagonal matrix
 \bea
\left[
\begin{array}{ccc}
\a & \d & \d \\ \d & \b & \d \\ \d & \d & \g \\
\end{array} \right]
 \eea
has rank 1 if and only if $\a=\b=\g=\d$. \epf

\subsection{Intersection of $\cP^4$ and $\S_{2,2}$}
Projective geometry has quite different properties from those
of the affine geometry. For instance, in the complex affine plane,
$\bC^2$, there exist distinct straight lines which are parallel
(and so do not meet). However, in the complex projective plane,
$\cP^2$, any two distinct lines meet at exactly one point.
More generally, any two projective varieties $X$ and $Y$ in
$\cP^n$ must meet if $\dim X+\dim Y\ge n$.

Let $V$ be a 5-dimensional subspace of the $3\times3$ system
$\cH=\cH_A\ox\cH_B$ and $\cP^4$ the projective 4-dimensional
subspace associated to $V$. Since the complex dimensions of $\cP^4$
and $\Sigma_{2,2}$, namely 4 and 4, add up to the dimension 8 of the
ambient projective space $\cP^8$, these two varieties must have
nonempty intersection, i.e., $V$ always contains at least one
product state, see e.g. \cite[Proposition 11.4]{h92}.
On the other hand $V$ may contain infinitely many
product states, i.e., the intersection of this $\cP^4$ and the Segre
variety $\Sigma_{2,2}$, may have positive dimension. Let us assume
that $V$ contains only finitely many product states which we treat
as points in $\cP^8$, say $P_i$, $i=1,\ldots,k$. This is often
expressed by saying that in this case the varieties $\cP^4$ and
$\Sigma_{2,2}$ intersect {\em properly}. Since $\cP^4$ and
$\Sigma_{2,2}$ have degrees 1 and 6 (see \cite[p. 233]{h92}),
respectively, B\'{e}zout Theorem tells us that $1\le k\le6$. More
precisely, we have $\m_1+\cdots+\m_k=6$ where $\m_i$, a positive
integer, is the intersection multiplicity of the point $P_i$. See
e.g. \cite{Gib,h92} for more details about these multiplicities
and the B\'{e}zout Theorem.

We next consider an arbitrary quintuple $(\ket{\psi_k})_{k=0}^4$ of
product vectors in general position and the 5-dimensional subspace
$V$ that they span. We would like to determine whether $V$ contains
any additional product states. The following proposition gives the
answer. Since the invariants determine uniquely (up to
BP-equivalence) the quintuple of points on $\Sigma_{2,2}$ in general
position, it is possible to analyze and answer the above question
solely in terms of the invariants.

 \bpp \label{prop:5generalposition}
Let $(P_k=\ket{\psi_k}=\ket{\phi_k}\ox\ket{\chi_k})_{k=0}^4$ be a
quintuple of product states in general position, and let $J_i^A$ and
$J_i^B$ $(i=1,2,3)$ be its invariants. Denote by $V$ the subspace
spanned by the $\ket{\psi_k}$ and by $\cP^4$ the associated
projective space. We consider the five equations:
\bea
\label{pet-jedn-3} && J_i^A=J_i^B, \quad (i=1,2,3), \\
\label{pet-jedn-4}
&& J_2^A(1-J_1^A)(1-J_2^B)=J_2^B(1-J_1^B)(1-J_2^A), \\
&& J_1^A(1-J_2^A)(1-J_1^B)=J_1^B(1-J_2^B)(1-J_1^A).
\label{pet-jedn-5}
\eea

(a) If any two of these equations hold, then all of them do. In that
case $V$ contains infinitely many product states. Moreover, any
state $\r$ with $\ker\r=V$ must be NPT.

(b) If exactly one of the above equations holds, then $V$ contains
no additional product states.

(c) If none of the above five equations holds, then $V$ contains
exactly one additional product state, say $\ket{\psi} =
\ket{c,z}=[c_iz_j]$. The vectors $\ket{c}$ and $\ket{z}$ are given
(up to a scalar factor) by the formulae \bea \label{c-formule}
\ket{c} = A^{-1} \left[ \begin{array}{c}
(1-J_1^B)/(J_1^B-J_1^A) \\
J_2^A(1-J_2^B)/(J_2^B-J_2^A) \\
(1-J_3^B)/J_1^A(J_3^B-J_3^A)
\end{array} \right],
\eea \bea \label{z-formule} \ket{z} = B^{-1} \left[ \begin{array}{c}
(1-J_1^A)/(J_1^B-J_1^A) \\
J_2^B(1-J_2^A)/(J_2^B-J_2^A) \\
(1-J_3^A)/J_1^B(J_3^B-J_3^A)
\end{array} \right],
 \eea
where $A$ and $B$ are the matrices which transform  the quadruples
$(\ket{\phi_k})_{k=0}^3$ and $(\ket{\chi_k})_{k=0}^3$ to the
canonical form, respectively. Moreover the six product states
$\ket{\ps_k}$, $k=0,\ldots,4$ and $\ket{\ps}$ are in general
position.
 \epp
 \bpf
We begin by transforming the quintuples $(\ket{\phi_k})_{k=0}^4$ and
$(\ket{\chi_k})_{k=0}^4$ by matrices $A$ and $B$, respectively.
Thus we may assume that the quadruples $(\ket{\phi_k})_{k=0}^3$ and
$(\ket{\chi_k})_{k=0}^3$ are in the canonical form, i.e., \bea
\label{spec-sl}
&& \ket{\psi_k}\propto\ket{kk}\quad (k=0,1,2), \\
&& \ket{\psi_3}=\sum_{i,j=0}^2 \ket{ij}, \\
&& \ket{\psi_4}=\ket{b,y}=[b_iy_j]. \eea Since the $P_k$ are in
general position, all components $b_i$ and $y_i$ are nonzero, and
$b_i\ne b_j$ and $y_i\ne y_j$ for $i\ne j$. A computation shows that
\bea
&& J_1^A=b_2/b_3,~J_2^A=b_3/b_1,~J_3^A=b_1/b_2; \\
&& J_1^B=y_2/y_3,~J_2^B=y_3/y_1,~J_3^B=y_1/y_2. \eea Thus we have
$J_i^A\ne1$ and $J_i^B\ne1$ for $i=1,2,3$, and the equations
(\ref{pet-jedn-4}) and (\ref{pet-jedn-5}) can be written as \bea
 b_1y_2+b_2y_3+b_3y_1 &=& b_1y_3+b_2y_1+b_3y_2, \\
\frac{1}{b_1y_2}+\frac{1}{b_2y_3}+\frac{1}{b_3y_1} &=&
\frac{1}{b_1y_3}+\frac{1}{b_2y_1}+\frac{1}{b_3y_2}, \eea
respectively.

We consider first the case (a). If two of the equations
(\ref{pet-jedn-3}) hold, so does the third because of the identity
$J_1J_2J_3=1$. Clearly, the remaining two equations also hold. The
other cases can be treated similarly. Let us just consider the
hardest case where the two equations displayed above hold. By
solving the first for $y_3$ and substituting into the second, we
obtain that
 \bea
\frac{(y_1-y_2)(b_1-b_3)(b_2-b_3)(b_1y_2-b_2y_1)}
{b_1b_2b_3y_1y_2(b_1y_2-b_2y_1+b_3y_1-b_3y_2)} = 0.
 \eea
Note that $b_1y_2-b_2y_1+b_3y_1-b_3y_2\ne0$ because $y_3\ne0$. Hence
we conclude that $b_1y_2-b_2y_1=0$, i.e., $J_3^A=J_3^B$. This means
that this case reduces to one of the other cases, and the first
assertion of (a) is proved.

We now assume that all five equations hold. We can further assume
that $b_1=y_1=1$ and, consequently, $b_2=y_2$ and $b_3=y_3$. Thus
$V$ consists of all symmetric matrices
 \bea
 \label{ea:symmetricmatrix}
\left[ \begin{array}{ccc}
u & \a+\b b_2 & \a+\b b_3 \\
\a+\b b_2 & v & \a+\b b_2b_3 \\
\a+\b b_3 & \a+\b b_2b_3 & w
\end{array} \right], \quad u,v,w,\a,\b\in\bC.
 \eea
By specializing the diagonal entries
 \bea
u = \frac{(\a+\b b_2)(\a+\b b_3)}{\a+\b b_2b_3}, \quad v =
\frac{(\a+\b b_2)(\a+\b b_2b_3)}{\a+\b b_3}, \quad w = \frac{(\a+\b
b_3)(\a+\b b_2b_3)}{\a+\b b_2}
 \eea
in the above matrix (\ref{ea:symmetricmatrix}), we obtain a family
of non-normalized product states depending on two complex parameters
$\a$ and $\b$. Since it is contained in $V$, the second assertion of
(a) is proved.

Let $\r$ be a non-normalized state with $\ker\r=V$. Its range,
$V^\perp$, is spanned by $\ket{01}-\ket{10}$, $\ket{12}-\ket{21}$,
$\ket{20}-\ket{02}$ and
$\ket{\ph}:=b(\ket{01}+\ket{10})-(1+b)(\ket{12}+\ket{21})+(\ket{20}+\ket{02})$,
where $b^*=-b_3(1-b_2)/b_2(1-b_3)$. Without loss of generality, we
can write $\r=\sum_{i=0}^3 \proj{\theta_i}$, where $\ket{\theta_i}$
are linearly independent non-normalized pure states
 \bea
\ket{\theta_0} &=& \ket{01}-\ket{10}, \\
\ket{\theta_1} &=& a_0(\ket{01}-\ket{10})+
    a_1(\ket{12}-\ket{21}), \\
\ket{\theta_2} &=& a_2(\ket{01}-\ket{10})+
    a_3(\ket{12}-\ket{21})+a_4(\ket{20}-\ket{02}), \\
\ket{\theta_3} &=& a_5(\ket{01}-\ket{10})+
    a_6(\ket{12}-\ket{21})+a_7(\ket{20}-\ket{02})+
    x\ket{\ph},
 \eea
with $a_1,a_4,x>0$. Assume that $\s:=\r^\G\ge0$. Since
$\ket{00}\in\ker\r$, the first diagonal entry of the matrix $\s$ is
0. Consequently, its first row must vanish. Hence we obtain that
 \bea
 \label{ea:row1}
a_4^2+(x-a_7)(x+a_7^*) &=& 0,  \\
a_2a_4+(bx+a_5)(x+a_7)^* &=& 0, \\
a_2^*a_4+(x-a_7)(bx-a_5)^* &=& 0, \\
 \label{ea:row4}
-1-|a_0|^2-|a_2|^2+(bx+a_5)(bx-a_5)^* &=& 0.
 \eea
From Eq. (\ref{ea:row1}) we deduce that $a_7$ is real and
$x^2=a_4^2+a_7^2$. From the next two we deduce that $a_5=-ba_7$ and
$a_2=b(a_7^2-x^2)/a_4$, and then Eq. (\ref{ea:row4}) gives the
contradiction $-1-|a_0|^2=0$.

Hence our assumption that $\s\ge0$ must be false, i.e., $\r$ must be
NPT. Thus all three assertions of (a) are proved.

Next we consider the case (b). Assume that one of the Eqs.
(\ref{pet-jedn-3}) holds. Say, $J_3^A=J_3^B$, i.e.,
$b_1y_2-b_2y_1=0$. Then we can also assume that $b_1=y_1=1$, and so
$b_2=y_2$. Hence $V$ consists of all matrices
 \bea
X=\left[
\begin{array}{ccc}
u & \a+\b b_2 & \a+\b y_3 \\
\a+\b b_2 & v & \a+\b b_2y_3 \\
\a+\b b_3 & \a+\b b_2b_3 & w
\end{array} \right], \quad u,v,w,\a,\b\in\bC.
 \eea
Assume that such a matrix $X$ has rank 1. Let us also assume that
$\a\b\ne0$. Then we can assume that $\b=1$. The following equations
must hold
\begin{eqnarray}
&& uv=(\a+b_2)^2, \\
&& u(\a+b_2b_3)=(\a+b_2)(\a+b_3), \\
&& u(\a+b_2y_3)=(\a+b_2)(\a+y_3).
\end{eqnarray}
From the last two equations we obtain that $u(\a+b_2b_3)(\a+y_3)=
u(\a+b_2y_3)(\a+b_3)$, i.e., $\a(b_2-1)(b_3-y_3)u=0$. As we deal
with case (b) and we assumed that $J_3^A=J_3^B$ and $b_1=y_1=1$, we
must have $J_2^A \ne J_2^B$, i.e., $b_3\ne y_3$. Hence, we deduce
that $u=0$, and so $\a=-b_2$. Since $X$ has rank 1 and $b_2\ne b_3$,
we deduce that the $(2,3)$ entry of $X$ must vanish. This gives the
contradiction $b_2(y_3-1)=0$. Consequently, we must have $\a\b=0$.
Then Lemma \ref{le:generic} implies that $X\propto\ket{\psi_k}$ for
some $k=0,\ldots,4$.

The case when Eq. (\ref{pet-jedn-4}) holds can be reduced to the
above case by switching the states $\ket{\psi_2}$ and
$\ket{\psi_3}$. After this transposition, the new values of the
invariants ${J'_i}^A$ are given by the formulae:
 \bea
{J'_1}^A=1-J_1^A,\quad {J'_2}^A=\frac{J_2^A}{J_2^A-1}, \quad
{J'_3}^A=\frac{J_2^A-1}{J_2^A(1-J_1^A)},
 \eea
and the same formulae are valid for ${J'_i}^B$. It remains to
observe that the equality ${J'_3}^A={J'_3}^B$ is the same as
(\ref{pet-jedn-4}).

The case when Eq. (\ref{pet-jedn-5}) holds can be reduced to the
previous case by switching the states $\ket{\psi_3}$ and
$\ket{\psi_4}$. After this transposition, the new values of the
invariants ${J'_i}^A$ are just the reciprocals of the old invariants
$J_i^A$, and similarly for ${J'_i}^B$. Then Eq. (\ref{pet-jedn-4})
is replaced by Eq. (\ref{pet-jedn-5}).
This completes the proof of (b).

It remains to prove (c). This is a generic case; the five subcases
of (b) can be intuitively viewed as the limiting cases of (c) which
occur when the state $\ket{\psi}$ (which we are going to construct)
becomes equal to one of the given five states so that its
multiplicity increases from 1 to 2.

Since $\ket{\psi}\in V$, we have \bea \ket{\psi}=\l \ket{\psi_0}+\m
\ket{\psi_1}+\n \ket{\psi_2}+ \a \ket{\psi_3}+\b \ket{\psi_4}. \eea
Lemma \ref{le:generic} implies that all coefficients must be
nonzero. The six off-diagonal entries of these matrices give the
following system of equations \bea && c_1z_2 = \a +\b b_1y_2, \quad
c_1z_3 = \a +\b b_1y_3,  \quad
c_2z_3 = \a +\b b_2y_3,  \\
&& c_2z_1 = \a +\b b_2y_1,  \quad c_3z_1 = \a +\b b_3y_1,  \quad
c_3z_2 = \a +\b b_3y_2. \label{sistem} \eea By eliminating $c_1$
from the first two equations, and $c_2$ and $c_3$ from the last
four, we obtain a system of linear equations in the unknowns $z_i$:
\bea \label{3-jed}
(\a +\b b_1y_3)z_2-(\a +\b b_1y_2)z_3 &=& 0, \\
-(\a +\b b_2y_3)z_1+(\a +\b b_2y_1)z_3 &=& 0, \\
(\a +\b b_3y_2)z_1-(\a +\b b_3y_1)z_2 &=& 0. \eea Since this system
has a nontrivial solution, its determinant must vanish:
 \bea
\left| \begin{array}{ccc}
0 & \a +\b b_1y_3 & -(\a +\b b_1y_2) \\
-(\a +\b b_2y_3) & 0 & \a +\b b_2y_1 \\
\a +\b b_3y_2 & -(\a +\b b_3y_1) & 0
\end{array} \right| = 0.
 \eea
This gives the equation
 \bea
(\a +\b b_1y_2) (\a +\b b_2y_3) (\a +\b b_3y_1)
 =(\a +\b b_1y_3) (\a +\b b_2y_1) (\a +\b b_3y_2).
 \eea
After expanding both sides, the terms with $\a^3$ and $\b^3$ cancel
and after dividing both sides by $\a\b$, we find that \bea
\label{ab-formule} \left[ \begin{array}{c} \a \\ \b \end{array}
\right] &\propto& \left[ \begin{array}{c}
b_2b_3y_1(y_3-y_2)+b_1b_3y_2(y_1-y_3)+b_1b_2y_3(y_2-y_1) \\
b_1(y_3-y_2)+b_2(y_1-y_3)+b_3(y_2-y_1)
\end{array} \right].
\eea We can replace the proportionality sign with the equality
because $\ket{c}$ and $\ket{z}$ are determined only up to a scalar
factor.

The system of linear equations (\ref{3-jed}) for the $z_i$ has the
unique solution up to a scalar factor: \bea \ket{z} \propto \left[
\begin{array}{c}
(\a +\b b_2y_1) (\a +\b b_3y_1) \\
 (\a +\b b_2y_1) (\a +\b b_3y_2) \\
(\a +\b b_2y_3) (\a +\b b_3y_1)
\end{array} \right],
\eea where $\a$ and $\b$ are given by Eq. (\ref{ab-formule}). After
substituting the expressions for $\a$ and $\b$ and cancelling two
factors, we obtain the formulae (\ref{z-formule}). From
(\ref{sistem}) we have \bea  \ket{c} &\propto& \left[
\begin{array}{c}
z_1z_3 (\a +\b b_1y_2) \\
z_1z_2 (\a +\b b_2y_3) \\
z_2z_3 (\a +\b b_3y_1)
\end{array} \right]
\eea and, by using (\ref{ab-formule}), we obtain (\ref{c-formule}).

Finally by using the above expressions, we can verify that the six
product states $\ket{\ps_k},k=0,\cdots,4$ and $\ket{\ps}$ are indeed
in general position. This completes the proof.
 \epf

We remark that, by B\'{e}zout Theorem, in the case (b) exactly one
of the 5 intersection points of $\cP^4\cap\S_{2,2}$ must have
multiplicity 2 and the other multiplicity 1. If $J_i^A=J_i^B$ holds,
then the point with multiplicity 2 is $P_{i-1}$. If Eq.
(\ref{pet-jedn-4}) [Eq. (\ref{pet-jedn-5})] holds, then this is the
point $P_3$ [$P_4$].

The case (c) is of special interest and we single it out in the
following definition.

\bd A quintuple of product states $(\ket{\psi_i})_{i=0}^4$ is {\em
regular} if it is in general position and the 5-dimensional subspace
spanned by the $\ket{\psi_i}$ contains exactly one additional product
state (up to scalar multiple). \ed

As an immediate consequence of the above proposition, we observe
that in the case (c) the map sending
$(\ket{b},\ket{y})\to(\ket{c},\ket{z})$ is involutory, i.e., it also
sends $(\ket{c},\ket{z})\to(\ket{b},\ket{y})$. As another
consequence, we have \bcr Let $(\ket{\psi_i})_{i=0}^4$ be a regular
quintuple of product states and $J_i^A,J_i^B$ its invariants. Denote
by $\ket{\psi}$ the unique additional product state in the subspace
spanned by the $\ket{\psi_i}$. Then the invariants
${J'_i}^A,{J'_i}^B$ of the quintuple
$(\ket{\psi_0},\ket{\psi_1},\ket{\psi_2},\ket{\psi_3},\ket{\psi})$
are given by the formulae \bea  &&
{J'_1}^A=\frac{(1-J_2^B)(J_3^B-J_3^A)}
{J_3^A(1-J_3^B)(J_2^B-J_2^A)}, \quad
{J'_2}^A=\frac{(1-J_3^B)(J_1^B-J_1^A)}
{J_1^A(1-J_1^B)(J_3^B-J_3^A)}, \quad
{J'_3}^A=\frac{(1-J_1^B)(J_2^B-J_2^A)}
{J_2^A(1-J_2^B)(J_1^B-J_1^A)}, \\  &&
{J'_1}^B=\frac{(1-J_2^A)(J_3^B-J_3^A)}
{J_3^B(1-J_3^A)(J_2^B-J_2^A)}, \quad
{J'_2}^B=\frac{(1-J_3^A)(J_1^B-J_1^A)}
{J_1^B(1-J_1^A)(J_3^B-J_3^A)}, \quad
{J'_3}^B=\frac{(1-J_1^A)(J_2^B-J_2^A)}
{J_2^B(1-J_2^A)(J_1^B-J_1^A)}. \eea \ecr

All cases (a-c) of the above proposition may occur; for (b) and (c)
see the last two examples in the proof of Theorem
\ref{thm:IntersPatterns}. In particular, the proposition shows that
the number of product states in $V$ may be infinite or only 5 even
when we impose the condition that the 5 given product states are in
general position. Nevertheless, we will show later that the kernel
of a $3\times3$ PPTES of rank 4 always contains 6 product states,
and that they are in general position
(see Theorem \ref{thm:3x3rank4PPTES,6productstates} in
Sec. \ref{sec:product=3x3rank4PPT}).
Hence, if the kernel of a state $\r$ of rank 4 is of type (a) or (b) of Proposition \ref{prop:5generalposition}, then $\r$ must be NPT.
This may shed new light on the problem of entanglement distillation
of $3\times3$ NPT states of rank 4 \cite{cd11}.
We investigate further the properties
of 5-dimensional subspaces in the next section.

The results proved in this section may help to solve another
long-standing quantum-information problem, namely the state
transformation under SLOCC \cite{dvc00}. The problem is completely
solved for $2\times M \times N$ pure states \cite{cc06,cms10};
however it becomes exceedingly difficult when all three dimensions
are bigger than two, e.g., deciding the SLOCC-equivalence of two
$3\times3\times3$ states. Here we consider the transformation
between two $3\times3\times5$ states $\ket{\ps}, \ket{\ph} \in \cH_A
\ox \cH_B \ox \cH_C$ and assume that there are 5 product states
$\ket{a_i,b_i} \in \tr_C \proj{\ps}$ in general position. According
to the range criterion \cite{cc06}, $\ket{\ps}, \ket{\ph}$ are
interconvertible via ILOs $V_A,V_B,V_C$ if and only if $V_A\ox V_B
\ket{a_i,b_i} \in \cR (\ph_{AB})$ where $\ph_{AB}=\tr_C \proj{\ph}$.
By finding out the product states in $\cR (\ph_{AB})$, we can decide
the SLOCC-equivalence of $\ket{\ps}, \ket{\ph}$ via Proposition
\ref{prop:Trans-4-5} and \ref{prop:5generalposition}. Hence, the
SLOCC-equivalence of two $3\times3\times5$ states can be
operationally decided provided that 5 product states in their range
of bipartite reduced states are available. Furthermore we can expand
subspaces, such as the $3\times3\times3$ and $3\times3\times4$
subspaces by respectively adding 2 or 1 linearly independent
(product) states, to span the whole $3\times3\times5$ space. Thus we
may treat the SLOCC-equivalence of tripartite states of the former
subspaces similar to the latter, when we can build 5 product states
in the range of corresponding reduced states.

\section{\label{sec:prodstate=3x3dim5} Product states in 5-dimensional
subspaces}

As in the previous section, let us consider the intersection
$\cP^4\cap\Sigma_{2,2}$ and assume that it is proper. Our main
objective here is to investigate various possibilities for this
intersection and provide concrete examples for each case.

\subsection{Intersection patterns}
Recall that because $\cP^4\cap\Sigma_{2,2}$ is a finite set,
consisting say of $k$ points, we know that necessarily $k\le6$.
Denote by $\m_i$ the intersection multiplicity of the point $P_i$.
When arranged in decreasing order $\m_1\ge\m_2\ge\cdots\ge\m_k$,
they form a partition $(\mu_1,\ldots,\m_k)$ of the integer 6.
We shall refer to this partition as the {\em intersection pattern}
(see \cite[p. 182]{Gib}).
Altogether there are 11 such partitions and we shall first prove
that all of them occur as intersection patterns.

\bt \label{thm:IntersPatterns} All 11 partitions of 6 occur as
intersection patterns of $\cP^4\cap\Sigma_{2,2}$, where $\cP^4$ is a
complex 4-dimensional projective space and $\Sigma_{2,2}$ the Segre
variety. \et

\bpf We just have to provide examples of two qutrit states
$\r$ of rank 4 whose kernel, viewed as a complex projective
4-dimensional subspace, has the specified partition of 6 as
its intersection pattern with $\Sigma_{2,2}$.
We shall subdivide the list into 6 parts corresponding to
the number, say $k$, of product states contained in $\ker\r$.
The corresponding partitions of 6 are those
having exactly $k$ parts. For $k=6$ the examples will be provided by
the normalized projectors associated with the UPB
(see Theorem \ref{thm:maintheorem=PPT3x3rank4})
and also by generic separable states of rank 4
(see Lemma \ref{le:separablekernel}). However we
shall include a concrete example with $k=6$ in our list.

In each case we assume that $\r=\sum_{i=0}^3\proj{\psi_i}$ and give
the formulae for the pure states $\ket{\psi_i}$. We also list the
$k$ product states in the $\ker\r$ as well as their intersection
multiplicities $\m_i$. These multiplicities were computed by means
of the free software package {\em Singular} \cite{GP} for symbolic
computation in Commutative Algebra.

For $k=1$ we set
 \bea
\ket{\psi_0}=\ket{12}, \quad \ket{\psi_1}=\ket{21}, \quad
\ket{\psi_2}=\ket{01}-\ket{10}-\ket{22}, \quad
\ket{\psi_3}=\ket{02}+\ket{11}-\ket{20}.
 \eea The kernel is spanned by
the states $\ket{00}$, $\ket{01}+\ket{10}$, $\ket{01}+\ket{22}$,
$\ket{02}+\ket{20}$ and $\ket{11}+\ket{20}$. The first one is the
only product state in the kernel. Its multiplicity must be 6.

For $k=2$ we give an example for each of the patterns $(5,1)$,
$(4,2)$ and $(3,3)$. For the first example we set
 \bea
\ket{\psi_0}=\ket{12}, \quad \ket{\psi_1}=\ket{01}-\ket{20}, \quad
\ket{\psi_2}=\ket{02}-\ket{21}, \quad
\ket{\psi_3}=\ket{10}-\ket{22}.
 \eea
The kernel is spanned by $\ket{00}$, $\ket{11}$,
$\ket{01}+\ket{20}$, $\ket{02}+\ket{21}$ and $\ket{10}+\ket{22}$.
The first two of them are the only product states in the kernel.
Their respective multiplicities are 5 and 1.

For the second example we set
 \bea
\ket{\psi_0}=\ket{01}-\ket{12}, \quad
\ket{\psi_1}=\ket{10}-\ket{21}, \quad
\ket{\psi_2}=\ket{02}-\ket{20}, \quad \ket{\psi_3}=\ket{22}.
 \eea One
can readily verify that $\ker \r$ is spanned by $\ket{00}$,
$\ket{11}$, $\ket{01}+\ket{12}$, $\ket{10}+\ket{21}$ and
$\ket{02}+\ket{20}$, and that $\ket{00}$ and $\ket{11}$ are the only
product states in the kernel. Their multiplicities are 2 and 4,
respectively.

For the third example we set
 \bea
\ket{\psi_0}=\ket{02}, \quad \ket{\psi_1}=\ket{01}-\ket{10}, \quad
\ket{\psi_2}=\ket{11}-\ket{20}, \quad
\ket{\psi_3}=\ket{12}-\ket{21}.
 \eea The kernel is spanned by
$\ket{00}$, $\ket{22}$, $\ket{01}+\ket{10}$, $\ket{11}+\ket{20}$ and
$\ket{12}+\ket{21}$. The first two of them are the only product
states in the kernel. Each of the two multiplicities is 3.

For $k=3$ we give examples for each of the patterns $(4,1,1)$,
$(3,2,1)$ and $(2,2,2)$. For the first example we set
 \bea
\ket{\psi_0}=\ket{02}, \quad \ket{\psi_1}=\ket{20}, \quad
\ket{\psi_2}=\ket{01}-\ket{12}, \quad
\ket{\psi_3}=\ket{10}-\ket{21}.
 \eea The kernel is spanned by
$\ket{00}$, $\ket{11}$, $\ket{22}$, $\ket{01}+\ket{12}$ and
$\ket{10}+\ket{21}$. The first three of them are the only product
states in the kernel. Their respective multiplicities are 1,4,1.

For the second example we set
 \bea
\ket{\psi_0}=\ket{20}, \quad \ket{\psi_1}=\ket{01}-\ket{22}, \quad
\ket{\psi_2}=\ket{02}-\ket{12}, \quad
\ket{\psi_3}=\ket{10}-\ket{21}.
 \eea The kernel is spanned by
$\ket{00}$, $\ket{11}$, $\ket{02}+\ket{12}$, $\ket{01}+\ket{22}$ and
$\ket{10}+\ket{21}$. The first three of these pure states are the
only product states in the kernel. Their respective multiplicities
are 1,2 and 3.

For the third example we set
 \bea
\ket{\psi_0}=\ket{11}, \quad \ket{\psi_1}=\ket{02}-\ket{12}, \quad
\ket{\psi_2}=\ket{20}-\ket{21}, \quad
\ket{\psi_3}=\ket{01}+\ket{10}+\ket{22}.
 \eea The kernel is spanned by
$\ket{00}$, $\ket{02}+\ket{12}$, $\ket{20}+\ket{21}$,
$\ket{01}+\ket{22}$ and $\ket{10}+\ket{22}$. The first three of
these pure states are the only product states in the kernel. For
each of them the intersection multiplicity is 2.

For $k=4$ we give examples with intersection patterns $(3,1,1,1)$
and $(2,2,1,1)$. For the first example we set
 \bea
\ket{\psi_0}=\ket{01}, \quad
\ket{\psi_1}=\ket{02}-\ket{11}+\ket{20}, \quad
\ket{\psi_2}=\ket{10}-\ket{22}, \quad
\ket{\psi_3}=\ket{12}-\ket{21}.
 \eea The kernel is spanned by
$\ket{00}$, $\ket{02}+\ket{11}$, $\ket{11}+\ket{20}$,
$\ket{10}+\ket{22}$ and $\ket{12}+\ket{21}$. The first product state
in the kernel is $\ket{00}$ and the other three are given by rank
one matrices:
 \bea
\left[ \begin{array}{lll}0&0&0\\1&\z&\z^2\\\z&\z^2&1
\end{array} \right], \quad \z^3=1.
 \eea
For $\ket{00}$ the intersection multiplicity is 3, and for the other
three points it is 1.

For the second example we set
 \bea
\ket{\psi_0}=\ket{10}, \quad \ket{\psi_1}=\ket{01}-\ket{22}, \quad
\ket{\psi_2}=\ket{02}-\ket{12}, \quad
\ket{\psi_3}=\ket{20}-\ket{21}.
 \eea The kernel is spanned by
$\ket{00}$, $\ket{11}$, $\ket{02}+\ket{12}$, $\ket{20}+\ket{21}$ and
$\ket{01}+\ket{22}$ . The first four of these pure states are the
only product states in the kernel. Their intersection multiplicities
are 1,1,2 and 2, respectively.

For $k=5$ there is only one possible intersection pattern, namely
$(2,1,1,1,1)$. We set
 \bea
\ket{\psi_0}=\ket{01}-\ket{20}, \quad
\ket{\psi_1}=\ket{02}-\ket{11}, \quad
\ket{\psi_2}=\ket{10}-\ket{22}, \quad
\ket{\psi_3}=\ket{12}-\ket{21}.
 \eea The kernel is spanned by the states
$\ket{00}$, $\ket{01}+\ket{20}$, $\ket{02}+\ket{11}$,
$\ket{10}+\ket{22}$ and $\ket{12}+\ket{21}$ . The first product
state in the kernel is $\ket{00}$ and the other four are given by
rank one matrices:
 \bea
\left[
\begin{array}{rrr}1&1&1\\1&1&1\\1&1&1\end{array} \right], \quad
\left[ \begin{array}{rrr}1&1&-1\\-1&-1&1\\1&1&-1\end{array} \right],
\quad \left[
\begin{array}{rrr}i&-i&1\\-1&1&i\\-i&i&-1\end{array} \right], \quad
\left[ \begin{array}{rrr}-i&i&1\\-1&1&-i\\i&-i&-1\end{array}
\right],
 \eea
where $i$ is the imaginary unit. For $\ket{00}$ the intersection
multiplicity is 2, and for the remaining four product states each
multiplicity is 1.

For $k=6$ there is again only one possible intersection pattern,
namely $(1,1,1,1,1,1)$. We set
 \bea
\ket{\psi_0}=\ket{01}-\ket{12}, \quad
\ket{\psi_1}=\ket{02}-\ket{21}, \quad
\ket{\psi_2}=\ket{10}-\ket{22}, \quad
\ket{\psi_3}=\ket{11}-\ket{20}.
 \eea The kernel is spanned by the states
$\ket{00}$, $\ket{01}+\ket{20}$, $\ket{02}+\ket{11}$,
$\ket{10}+\ket{22}$ and $\ket{12}+\ket{21}$ . The first product
state in the kernel is $\ket{00}$ and the other five are given by
rank one matrices:
 \bea
\left[ \begin{array}{lll}1&\x^4&\x\\ \x^3&\x^2&\x^4\\
\x^2&\x&\x^3
\end{array} \right], \quad \x^5=1. \eea
Clearly, in this case each multiplicity must be 1.
 \epf
The above example for $k=5$ shows that a state of rank 4 whose range
contains no product state may fail to be a PPTES. Indeed, the kernel
of a PPTES of rank 4 contains exactly 6 product states
(see Theorem \ref{thm:3x3rank4PPTES,6productstates} below).

On the other hand the example that we chose for $k=6$ is
SLOCC-equivalent to the {\bf Pyramid} UPB in \cite{DiV03}. One way
to see this is simply to verify that the quintuple of product states
given by the above matrices for $\xi=\exp(2\pi ik/5)$, $k=0,\ldots,4$
and the UPB quintuple for the {\bf Pyramid} (see Eq.
(\ref{alfa-beta}) below) have the same invariants.
If we exchange the parties A,B and transform the  product states in
the kernel by the local operator $S\ox S$,
$S=\left[\begin{array}{cc}1&1\\i&-i\end{array}\right]
\oplus\left[\sqrt{1+\sqrt5}\right]$, we again obtain the {\bf Pyramid}.

Of course, the intersection of $\cP^4$ and $\Sigma_{2,2}$ does not
have to be proper, i.e., it may have positive dimension. For the
comparison with Theorem \ref{thm:IntersPatterns}, we provide a
scenario where there are infinitely many product states in the
kernel.

 \bl \label{le:3x3rank4,rankC=1} Let $\r$ be a $3\times 3$
state of rank 4 such that,  $\rank\bra{x}\r\ket{x}=1$ for some
$\ket{x}\in\cH_A$. Then $\ker\r$ contains infinitely many
product states.
 \el
 \bpf
Since the operator $\s=\bra{x}\r\ket{x}$ has rank one, it suffices
to observe that the 2-dimensional subspace $\ket{x}\ox\ker(\s)$ is
contained in the kernel of $\r$.
 \epf

\subsection{Rank-4 PPTES with no product state in the range}
In this subsection we consider a related problem of
describing the product states in the kernel of states of rank 4 having
no product state in the range. The set of such states properly
contains all $3\times3$ PPTES of rank 4, as one will see later in
Theorem \ref{thm:maintheorem=PPT3x3rank4}. It is thus
important to have a general understanding of this set.
 \bl
 \label{le:range=1prodstate}
Let $\r$ be a $3\times3$ state of rank 4. Then $\cR(\r)$ contains at
least one product state when its kernel contains either

(a) two linearly independent product states $\ket{a,b}$ and
$\ket{c,d}$ with $\ket{a}=\ket{c}$ or $\ket{b}=\ket{d}$, or

(b) three linearly independent product states
 $\ket{u_i,v_i}$, $i=1,2,3$ such that the $\ket{u_i}$ or
the $\ket{v_i}$ are linearly dependent.
 \el
 \bpf
If (a) holds, say $\ket{a}=\ket{c}$, then the 7-dimensional space
$\lin\{\ket{a,b},\ket{a,d}\}^\perp$ contains $\cR(\r)$
and the 6-dimensional subspace $V=\ket{a}^\perp \ox \cH_B$.
Since $V\cap\cR(\r)$ has dimension $\ge3$, it must contain a product
state.

If (b) holds, say the $\ket{u_i}$ are linearly dependent, then the
6-dimensional space $\{\ket{u_i,v_i}:i=1,2,3\}^\perp$ contains
$\cR(\r)$ and the 3-dimensional subspace
$V=\{\ket{u_i},i=1,2,3\}^\perp \ox \cH_B$. Hence, the subspace
$V\cap\cR(\r)$ has dimension $\ge1$. Since each nonzero vector in
$V$ is a product state the assertion follows. This completes the
proof.
 \epf

Note that the converse does not hold; i.e., both (a) and (b) may
fail even though $\cR(\r)$ contains a product state
(see the first example for $k=3$). Nevertheless, this
result will be strengthened in the case of PPTES
in Sec. \ref{sec:product=3x3rank4PPT}.

On the other hand there may exist only 3 product states in general
position in a 5-dimensional kernel, when there is no product state
in the range of $\r$. We consider 5 states in $\ker\r$ represented
by the matrices $C_i,i=0,1,2,3,4$
 \bea \label{ea:3prod=kernel,0prod=range}
\left[ \begin{array}{ccc}
1 & 0 & 0 \\
0 & 0 & 0 \\
0 & 0 & 0
\end{array}\right],\quad
\left[\begin{array}{ccc}
0 & 0 & 0 \\
0 & 1 & 0 \\
0 & 0 & 0
\end{array}\right],\quad
\left[\begin{array}{ccc}
0 & 0 & 0 \\
0 & 0 & 0 \\
0 & 0 & 1
\end{array}\right],\quad
\left[\begin{array}{ccc}
0 & 0 & 1 \\
0 & 0 & 1 \\
1 & 1 & 0
\end{array}\right],\quad
\left[\begin{array}{ccc}
0 & a & b \\
a & 0 & -1+b \\
1 & 0 & 0
\end{array}\right],
 \eea
respectively, where $a\ne0, b\ne1$. Up to ILOs, this is also the
generic expressions of basis in a 5-dimensional kernel where there
are only 3 product states in general position and there are no
product states in the range of $\r$.
 \bl
 \label{le:3prod=kernel,0prod=range}
For any $3\times3$ state $\r$ of rank 4, whose kernel is spanned by the
pure states (\ref{ea:3prod=kernel,0prod=range}), $\cR(\r)$ contains
no product state while $\ker\r$ contains only 3 product states and
they are in general position.
 \el
 \bpf
First we show that there is no product state $\ket{a,b}\in\cR(\r)$.
Because the first three product states in the kernel are
$\ket{00},\ket{11},\ket{22},$ we have that
$\ket{a,b}=\ket{i}(x_i\ket{(i+1)\mod3}+y_i\ket{(i+2)\mod3}),$ or
$(x_i\ket{(i+1)\mod3}+y_i\ket{(i+2)\mod3})\ket{i}$, $i=0,1,2$.
Since one of them is orthogonal to the latter two states
$C_3,C_4$ in Eq. (\ref{ea:3prod=kernel,0prod=range}),
there are two parallel rows or columns in the same position of
$C_3,C_4$. This is evidently impossible,
so there is no product state in $\cR(\r)$.

Second we show that $\ket{00},\ket{11},\ket{22}$ are the only three
product states in the kernel. It is easy to see that we must use
$C_4$. We compute the linear combination of $C_i,i=0,1,2,3,4$ such
that
 \bea C:=u C_0 + v C_1 + w C_2 + x C_3 + C_4 =
 \left[\begin{array}{ccc}
u &   a & b+x \\
a &   v & -1+b+x \\
1+x & x & w
\end{array}\right],
 \eea
which has rank 1 when it is a product state. Evidently $x\ne-1,0$.
Then we can deduce that $v=\frac{a x}{1+x}$, which leads to $\det
\left[\begin{array}{cc}
a & b+x \\
v & -1+b+x \\
\end{array}\right] \ne0$. Hence $C$ cannot be a product state $\forall
u,v,w,x$. This completes the proof.
 \epf

Furthermore we consider 5 states in $\ker\r$ represented by the
matrices $C_i,i=0,1,2,3,4$
 \bea \label{ea:2prod=kernel,0prod=range}
\left[ \begin{array}{ccc}
1 & 0 & 0 \\
0 & 0 & 0 \\
0 & 0 & 0
\end{array}\right],\quad
\left[\begin{array}{ccc}
0 & 0 & 0 \\
0 & 1 & 0 \\
0 & 0 & 0
\end{array}\right],\quad
\left[\begin{array}{ccc}
0 & 1 & 1 \\
0 & 0 & 1 \\
0 & 0 & 1
\end{array}\right],\quad
\left[\begin{array}{ccc}
0 & 1 & 1 \\
0 & 0 & 0 \\
1 & -2-c & -1-c
\end{array}\right],\quad
\left[\begin{array}{ccc}
0 & c-\frac{8}{2+c} & c \\
1 & 0 & 0 \\
0 & \frac{4c}{2+c} & c
\end{array}\right],
 \eea
respectively, where the real $c\ne -2,0$. Then we have
 \bl
 \label{le:2prod=kernel,0prod=range}
For any $3\times3$ state $\r$ of rank 4, whose kernel is spanned by
the pure states (\ref{ea:2prod=kernel,0prod=range}), $\cR(\r)$
contains no product states while $\ker\r$ contains
only 2 product states (up to scalar multiple) and these two
states are in general position.
 \el
 \bpf
The proof is similar to that for
Lemma \ref{le:3prod=kernel,0prod=range}.
First we show that there is no
product state $\ket{a,b}\in\cR(\r)$.
Because the first two product
states in the kernel are $\ket{00},\ket{11}$ we have that
$\ket{a,b}=(x_0\ket{0}+x_2\ket{2})(y_1\ket{1}+y_2\ket{2}),$ or
$(x_1\ket{1}+x_2\ket{2})(y_0\ket{0}+y_2\ket{2}),$ or $\ket{2,b}$ or
$\ket{a,2}$. One can check that none of them exists, so there is no
product state in $\cR(\r)$. Second we show that $\ket{00},\ket{11}$
are the only 2 product states in the kernel. To simplify the proof,
we notice that the the first 4 blocks cannot generate new product
states. So we must need $C_4$. Further we consider three cases for
the linear combination $C:=u C_0 + v C_1 + w C_2 + x C_3 + C_4 $,
namely $g=4c/(2+c)^2$, $g=1$ and the rest. In each of these cases,
one can easily show that $C$ cannot be a product state. This
completes the proof.
 \epf

We lack examples of states whose range contains no product
state, and whose kernel contains only 1 product state. This problem,
as well as Lemmas \ref{le:3prod=kernel,0prod=range} and
\ref{le:2prod=kernel,0prod=range}, is more relevant for the
characterization of NPT states, which is an essentially useful
quantum-information resource (for a recent paper see \cite{cd11}).
From the next section we will focus on the main topic of this paper,
namely the description of the $3\times3$ PPTES of rank 4 via
the UPB construction \cite{bdm99}.

\section{Characterization of equivalent $3\times3$ UPB} \label{sec:equivalentUPB}

Let us denote by $\cE_4$ the set of all PPTES of rank 4 in a
$3\ox 3$ system. Our main objective is to prove a conjecture which
was raised in \cite{lms10} and gives a full description of the set
$\cE_4$. This has close connection with the family of PPTES
constructed in \cite{DiV03} via UPBs.

\subsection{PPTES of rank 4 and UPB}
Let $\cU$ denote the set of all UPBs in $\cH=\cH_A\ox\cH_B$. We
denote by $\cU^\circ$ the set of all quintuples
$(\psi):=(\ket{\psi_k}=\ket{\phi_k}\ox\ket{\chi_k})_{k=0}^4$ of
(normalized) product states such that the set $\{\ket{\psi_k}:0\le
k\le4 \}$ is a UPB and the following orthogonality relations hold:
\bea \label{rel-ortog} &&
\braket{\phi_i}{\phi_{i+1}}=\braket{\chi_i}{\chi_{i+2}}=0, \eea
where the indexes are taken modulo 5. We have a natural projection
map $\cU^\circ\to\cU$ which associates to a quintuple
$(\psi)=(\ket{\psi_k})_{k=0}^4\in\cU^\circ$ the UPB
$\{\psi\}:=\{\ket{\psi_k}:0\le k\le4 \}\in\cU$. It was shown in
\cite{DiV03} that this map is onto. It is clearly 10-to-1 map
because the cyclic permutation of the $\ket{\psi_i}$ and also the
reflection which interchanges the indexes via the permutation
$(0)(14)(23)$ has no effect on the set $\{\psi\}$, and leaves
$\cU^\circ$ globally invariant.

There is a natural map $\Pi:\cU\to\cE_4$ which associates to
$\{\psi\}\in\cU$ the state $\Pi\{\psi\}:=(1/4)P$, where $P$ is the
orthogonal projector of rank 4 with $\ker P=\lin \{\psi\}$. The
following conjecture, which gives explicit description of $\cE_4$
was raised recently in \cite{lms10} and supported by vast numerical
evidence.

\bcj \label{conj:HipLMS} Every state $\r\in\cE_4$ is the normalization
of $A\ox B~ \Pi\{\psi\}~ A^\dag\ox B^\dag$ for some $(A,B)\in\GL$
and $\{\psi\}\in\cU$. \ecj

The proof of this conjecture will be given in Theorem
\ref{thm:maintheorem=PPT3x3rank4} of Section
\ref{sec:product=3x3rank4PPT}.

We fix o.n. bases $\{\ket{0}_A,\ket{1}_A,\ket{2}_A\}$ and
$\{\ket{0}_B,\ket{1}_B,\ket{2}_B\}$ of $\cH_A$ and $\cH_B$,
respectively. It was shown in \cite{DiV03} that every quintuple
$\ket{\psi_k}=\ket{\a_k}\ox\ket{\b_k}$, $k=0,\ldots,4$, in
$\cU^\circ$ is LU-equivalent to one in the following 6-parameter
family:
 \bea
 \label{alfa-beta}
&& \ket{\a_0}=\ket{0}_A,  \\
&& \ket{\a_1}=\ket{1}_A,  \\
&& \ket{\a_2}=\cos\t_A\ket{0}_A+\sin\t_A\ket{2}_A,  \\
&& \ket{\a_3}=\sin\g_A\sin\t_A\ket{0}_A-\sin\g_A
\cos\t_A\ket{2}_A+\cos\g_A e^{i\phi_A}\ket{1}_A,  \\
&& \ket{\a_4}=\frac{1}{N_A}\left( \sin\g_A\cos\t_A
e^{i\phi_A}\ket{1}_A+\cos\g_A\ket{2}_A
\right),  \\
&& \ket{\b_0}=\ket{1}_B,  \\
&& \ket{\b_1}=\sin\g_B\sin\t_B\ket{0}_B-
\sin\g_B\cos\t_B\ket{2}_B+\cos\g_B e^{i\phi_B}\ket{1}_B,  \\
&& \ket{\b_2}=\ket{0}_B,  \\
&& \ket{\b_3}=\cos\t_B\ket{0}_B+\sin\t_B\ket{2}_B,  \\
&& \ket{\b_4}=\frac{1}{N_B}\left( \sin\g_B\cos\t_B
e^{i\phi_B}\ket{1}_B+\cos\g_B\ket{2}_B \right).
 \eea The 6 real parameters are the angles: $\g_{A,B}$, $\t_{A,B}$
and $\phi_{A,B}$, and the normalization constants $N_{A,B}$ are
given by the formulae
 \bea
N_{A,B}=\sqrt{\cos^2\g_{A,B}+\sin^2\g_{A,B}\cos^2\t_{A,B}}.
 \eea The
first four angles are required to have nonzero sine and cosine,
while the angles $\phi_{A,B}$ may be arbitrary. It is not hard to
show that the parameter domain can be further restricted as in the
following lemma.
 \bl \label{le:domen} Every quintuple
$(\psi)=(\ket{\psi_k})_{k=0}^4\in\cU^\circ$ with
$\ket{\psi_k}=\ket{\a_k}\ox\ket{\b_k}$ is LU-equivalent to one
belonging to the family (\ref{alfa-beta}) such that the four angles
$\g_{A,B}$, $\t_{A,B}$ belong to the interval $(0,\pi/2)$. \el \bpf
We may assume that $\ket{\a_0}=\ket{0}_A$, $\ket{\a_1}=\ket{1}_A$,
$\ket{\b_0}=\ket{1}_B$ and $\ket{\b_2}=\ket{0}_B$. As $\ket{\a_2}$
is a linear combination of $\ket{0}_A$ and $\ket{2}_A$, we can
choose the overall phase of $\ket{\a_2}$ so that the coefficient of
$\ket{0}_A$ is positive. By applying a diagonal unitary matrix
$U_A=\diag(1,1,\xi)$, we can also assume that the coefficient of
$\ket{2}_A$ is positive. Thus
$\ket{\a_2}=\cos\t_A\ket{0}_A+\sin\t_A\ket{2}_A$ for some
$\t_A\in(0,\pi/2)$.

Next, $\ket{\a_3}$ is a linear combination of $\ket{1}_A$ and
$\sin\t_A\ket{0}_A-\cos\t_A\ket{2}_A$. We can choose the overall
phase of $\ket{\a_3}$ so that the coefficient of
$\sin\t_A\ket{0}_A-\cos\t_A\ket{2}_A$ is positive. Thus \bea
\ket{\a_3}=\sin\g_A\sin\t_A\ket{0}_A -\sin\g_A\cos\t_A\ket{2}_A +
\cos\g_A e^{i\phi_A}\ket{1}_A \eea for some $\g_A\in(0,\pi/2)$ and
some angle $\phi_A$.

Finally, $\ket{\a_4}$ is a linear combination of $\ket{1}_A$ and
$\ket{2}_A$. We can choose the overall phase of $\ket{\a_2}$ so that
the coefficient of $\ket{2}_A$ is positive. Since $\ket{\a_4}$ is
orthogonal to $\ket{\a_3}$, we have \bea \ket{\a_4}=\frac{1}{N_A}
\left( \sin\g_A\cos\t_A e^{i\phi_A} \ket{1}_A +\cos\g_A\ket{2}_A
\right) \eea with $\g_A\in(0,\pi/2)$ and the positive normalization
constant $N_A$.

The same arguments can be used on Bob's side. \epf

We shall give just one example, namely the UPB quintuple known as
{\bf Tiles}. Its parameters, as given in \cite[p. 394]{DiV03}, are
$\phi_{A,B}=0$, $\t_{A,B}=\g_{A,B}=3\pi/4$. This quintuple is
LU-equivalent to the one given by parameters $\phi_{A,B}=\pi$,
$\t_{A,B}=\g_{A,B}=\pi/4$. The local unitary transformation that we
can use in this case fixes the vectors $\ket{0}_A, \ket{1}_A$ and
$\ket{0}_B, \ket{1}_B$, and sends $\ket{2}_A$ and $\ket{2}_B$ to
their negatives.

Let $\cF$ be the subfamily of the family (\ref{alfa-beta}) obtained
by restricting the domain of parameters so that the four angles
$\g_{A,B}$, $\t_{A,B}$ belong to the interval $(0,\pi/2)$ while the
angles $\phi_{A,B}$ belong to $(-\pi,\pi]$. Since the domain of
parameters is connected, the family $\cF$ is also connected. By
Lemma \ref{le:domen} we have $\cU^\circ=\Un(3) \times \Un(3) \cdot
\cF$, and so the set $\cU^\circ$ is connected too.

\bl \label{prop:GL->U} Assume that $A\ox B (\psi)=(\psi')$ where
$(A,B)\in\GL$ and the quintuples $(\psi)=(\ket{\psi_k})_{k=0}^4$ and
$(\psi')=(\ket{\psi'_k})_{k=0}^4$ belong to $\cU^\circ$. Then there
is a positive constant $c$ such that $cA$ and $c^{-1}B$ are unitary.
In particular, if $(\psi)$ and $(\psi')$ are SLOCC-equivalent then
they are LU-equivalent.
 \el
 \bpf
Let us write \bea \ket{\psi_k}=\ket{\phi_k}\ox\ket{\chi_k}, \quad
\ket{\psi'_k}=\ket{\phi'_k}\ox\ket{\chi'_k}, \quad k=0,\ldots,4.
\eea By the hypothesis we have \bea A\ket{\phi_k}\ox
B\ket{\chi_k}=\ket{\phi'_k}\ox\ket{\chi'_k}, \quad k=0,\ldots,4.
\eea Hence, $A\ket{\phi_k}=a_k\ket{\phi'_k}$ and
$B\ket{\chi_k}=a_k^{-1}\ket{\chi'_k}$ for some scalars $a_k$. Since
$\braket{\phi'_k}{\phi'_{k+1}}=0$, we deduce that
$\bra{\phi_k}A^\dag A\ket{\phi_{k+1}}=0$. Consequently, $A^\dag A$
must map the plane spanned by $\ket{\phi_{k+1}}$ and
$\ket{\phi_{k-1}}$ onto itself. This clearly implies that $A^\dag A$
is a scalar matrix, i.e., there is a scalar $c>0$ such that $cA$ is
a unitary matrix. A similar argument shows that $c^{-1}B$ is also
unitary. \epf

Note that SLOCC-equivalence is different from the BP-equivalence
which does not require the identical global scalar for simultaneous
transformations $A \ox B \ket{\psi_k}=\ket{\psi'_k}, k=0,\cdots,4$.
By Lemma \ref{prop:GL->U}, two SLOCC-equivalent quintuples $(\psi)$
and $(\psi')$ in $\cF$ must be connected by local unitary operators
$U_A \ox U_B$.

 \bpp
If two quintuples $(\ket{\psi_k}),(\ket{\psi'_k})\in\cF$ are
SLOCC-equivalent, then $\ket{\psi_k}=\ket{\psi'_k}$ for
$k=0,\ldots,4$.
 \epp
\bpf As in Eqs. (\ref{alfa-beta}) we write
$\ket{\psi_k}=\ket{\a_k}\ox\ket{\b_k}$ and similarly let
$\ket{\psi'_k}=\ket{\a'_k}\ox\ket{\b'_k}$. Let
$(\g_{A,B},\t_{A,B},\phi_{A,B})$ and
$(\g'_{A,B},\t'_{A,B},\phi'_{A,B})$ be the parameters of the two
quintuples, respectively.

The SLOCC-equivalence implies that the quintuples
$(\ket{\a_k})_{k=0}^4$ and $(\ket{\a'_k})_{k=0}^4$ are projectively
equivalent. We point out that these quintuples are in general
position and so, by Proposition \ref{prop:5-Points} and
\ref{prop:Trans-4-5} they must have the same invariants $J_i^A$. By
using the expressions in Eq. (\ref{alfa-beta}) and the formulae Eq.
(\ref{J1-J2}), we find that these invariants for $(\ket{\psi_k})$
are given by \bea \label{UPB-inv}
\begin{array}{lll}
J_1^A= -\tan^2\g_A\cos^2\t_A, & J_2^A=\frac{1}{\cos^2\t_A}, &
J_3^A = -\cot^2\g_A,  \\
J_1^B=\frac{1}{\sin^2\t_B}, &
J_2^B=\frac{\sin^2\t_B}{1+\cos^2\t_B\tan^2\g_B}, &
J_3^B=1+\cos^2\t_B\tan^2\g_B.
\end{array}
\eea Similar formulae are valid for the quintuple $(\ket{\psi'_k})$.
The equalities $J_i^A=J_i^{A'}$ and $J_i^B=J_i^{B'}$ imply that \bea
\cos^2\xi_X=\cos^2\xi'_X, \quad \xi=\g,\t;~ X=A,B. \eea Since all
these angles belong to $(0,\pi/2)$, we conclude that
$\g'_{A,B}=\g_{A,B}$ and $\t'_{A,B}=\t_{A,B}$. It remains to prove
that also $\phi'_{A,B}=\phi_{A,B}$.

By Lemma \ref{prop:GL->U}, there exist unitary matrices $A=[a_{ij}]$
and $B=[b_{ij}]$ such that \bea \label{jednacine}  A\ket{\a_k} \ox
B\ket{\b_k} - \ket{\a'_k} \ox \ket{\b'_k}=0, \quad k=0,\ldots,4.
\eea

We view each of these equations as a matrix equation. For $k=0$ it
gives $a_{11}b_{22}=1$ and $a_{21}=a_{31}=b_{12}=b_{32}=0$. Thus we
may assume that $a_{11}=b_{22}=1$ and, since $A$ and $B$ are
unitary, we must also have $a_{12}=a_{13}=b_{21}=b_{23}=0$.

For $k=2$, the matrix equation shows that $b_{11}=1$ and $a_{33}=1$.
We deduce that $a_{23}=a_{32}=b_{13}=b_{31}=0$.

Now, for $k=1$ the matrix equation implies that $a_{22}=b_{33}=1$.
This means that $A=B=I_3$ and so $\phi'_{A,B}=\phi_{A,B}$.
 \epf

\subsection{UPB symbols}
Let us say that a 5-dimensional subspace $W\subseteq\cH$ is of {\em
UPB type} if it is BP-equivalent to a subspace spanned by a UPB. We
can characterize the UPB-type subspaces by using invariants. For
this purpose we attach a 6-letter symbol made up of letters N, P and
p to each quintuple of product states in general position having all
invariants real.

Let us denote the open intervals $(-\infty,0)$, $(0,1)$,
$(1,+\infty)$ by the letters $N$, $p$, $P$, respectively. ($N$ is
for ``negative'', $p$ for ``positive and small'' and $P$ for
``positive and large''.) Let $(\psi):=(\ket{\psi_k})_{k=0}^4$, be
any quintuple of product states in general position. For
convenience, we shall say that $(\psi)$ is {\em real} if all of its
invariants are real numbers. Since the invariants do not take values
0 and 1, if $(\psi)$ is real its invariants must take the values in
one of the intervals N,p,P. For real $(\psi)$ we define its {\em
symbol} to be the 6-letter sequence obtained by replacing each of
its invariants $J_1^A,J_2^A,J_3^A,J_1^B,J_2^B,J_3^B$ by the interval
to which it belongs. For instance, each quintuple $(\psi)\in\cF$ has
$NPNPpP$ as its symbol because $J_1^A<0$, $J_2^A>1$, $J_3^A<0$,
$J_1^B>1$, $0<J_2^B<1$ and $J_3^B>1$. Altogether there are 144
symbols that arise in this manner. Each symbol can be broken into
two parts, A and B. The A [B] part consists of the first [last]
three letters. Because of the identity $J_1J_2J_3=1$, there are only
12 possibilites for each of the two parts: \bea NNP,~
NNp,~NPN,~NpN,~pNN,~pPP,~pPp,~ppP,~PNN,~PPp,~PpP,~Ppp. \eea We shall
refer to the 12 symbols in Table 1 as the {\em UPB symbols}.

\begin{center}
\begin{tabular}{ll|ll|ll|ll}
\multicolumn{8}{c} {Table 1: UPB symbols and associated
permutations} \\ \hline NNPPPp & (12)(34) & NNpppP & (12) &
NPNPpP & {\rm id} & NpNpPp & (34) \\
PNNpPP & (01)(34) & PPpNNP & (23) &
PpPNPN & (1243) & PpppNN & (02)(14) \\
pNNPpp & (01) & pPPPNN & (13) &
pPpNpN & (03)(24) & ppPNNp & (01)(23) \\
\hline
\end{tabular}
\end{center}

We can now prove the main result of this section.

\bt \label{thm:UPB-type} Let $W\subseteq\cH$ be a 5-dimensional
subspace containing exactly 6 product states (up to scalar factors)
and assume that these states are in general position. Denote by
$\Psi$ the collection of the 720 quintuples
$(\psi):=(\ket{\psi_k})_{k=0}^4$, selected from these 6 product
states.

(a) Assume $(\psi)\in\Psi$ is real and its symbol is UPB. If $\s$ is
the permutation from Table 1 associated to this symbol, then
$(\ket{\psi_{\s k}})_{k=0}^4$ has symbol $NPNPpP$.

(b) If some $(\psi)\in\Psi$ is real and its symbol is UPB, then $W$
has UPB type.

(c) Conversely, if $W$ has UPB type, then all $(\psi)\in\Psi$ are
real and their symbols are UPB. \et

\bpf The assertion (a) can be proved by straightforward
verification. We shall give details for one case which we shall need
later. Assume that $(\psi)$ has symbol $pNNPpp$. Since we work with
non-normalized states, without any loss of generality we may assume
that $(\psi)$ is given by the pair of matrices \bea  A=\left[
\begin{array}{ccccc} 1&0&0&1&1\\0&1&0&1&-a\\0&0&1&1&-b
\end{array} \right],\quad
B=\left[ \begin{array}{ccccc} 1&0&0&1&1\\0&1&0&1&c\\0&0&1&1&1/d
\end{array} \right].
\eea By computing the invariants of $(\psi)$ we obtain that \bea
J_1^A=a/b,~J_2^A=-b,~J_3^A=-1/a,~ J_1^B=cd,~J_2^B=1/d,~J_3^B=1/c,
\eea and so we must have $b>a>0$ and $c,d>1$. From Table 1 we find
that $\s=(01)$ in this case. Hence, the quintuple
$(\psi'):=(\ket{\psi_{\s k}})_{k=0}^4$ is given by the matrices $A'$
and $B'$ which are obtained from $A$ and $B$, respectively, by
switching the first two columns. For the invariants of $(\psi')$ we
obtain the formulae \bea {J'_1}^A=-1/b,~{J'_2}^A=b/a,~{J'_3}^A=-a,~
{J'_1}^B=d,~{J'_2}^B=1/cd,~{J'_3}^B=c. \eea Thus, the symbol of
$(\psi')$ is indeed $NPNPpP$.

Next we prove (b). By using (a) we may assume that $(\psi)$ has
symbol $NPNPpP$. Since $J_1^A J_2^AJ_3^A=1$ and $J_1^B J_2^B
J_3^B=1$, the equations (\ref{UPB-inv}) can be solved for the four
angles $\g_A,\theta_A,\g_B,\theta_B$. Now the assertion follows from
Propositions \ref{prop:5-Points} and \ref{prop:Trans-4-5}.

To prove (c) we observe that any pair of real matrices \bea U=\left[
\begin{array}{ccccc} 1&0&\a&\b&0\\0&1&0&1&\a\\0&0&\b&-\a&1
\end{array} \right],\quad
V=\left[ \begin{array}{ccccc}
1&\delta&0&0&\g\\0&1&1&\g&0\\0&-\g&0&1&\delta
\end{array} \right],
\eea with nonzero parameters $\a,\b,\g,\d$ defines a non-normalized
UPB given by the tensor products of the corresponding columns of $U$
and $V$. Moreover, any UPB is BP-equivalent to one of this form. The
invariants of the above UPB are
 \bea
\left[ -\a^2, \frac{\a^2+\b^2}{\a^2},-\frac{1}{\a^2+\b^2},
1+\g^2,\frac{\d^2}{(1+\g^2)(\g^2+\d^2)},\frac{\g^2+\d^2}{\d^2}
\right].
 \eea
Evidently, this quintuple has the symbol $NPNPpP$. By using these
invariants and the formulae Eqs. (\ref{c-formule}) and
(\ref{z-formule}) from Proposition \ref{prop:5generalposition} we
find the sixth product state in $W$ and extend the matrices $U$ and
$V$ to $3\times6$ matrices \bea  \tilde{U} &=& \left[
\begin{array}{cccccc} 1&0&\a&\b&0&
\a[(1+\a^2)(1+\g^2+\d^2)+\b^2(\g^2+\d^2)]/
[\b(1+\a^2+\g^2)] \\
0&1&0&1&\a& \a(1+\g^2+\d^2)[\d^2+(\a^2+\b^2)(\g^2+\d^2)]/
[\a^2\g^2+(\g^2+\d^2)(\b^2+\g^2(\a^2+\b^2))] \\
0&0&\b&-\a&1&1
\end{array} \right], \\
\tilde{V} &=& \left[ \begin{array}{cccccc} 1&\delta&0&0&\g&
\g[\a^2(1+\a^2+\b^2)(1+\g^2+\d^2)+\b^2(1+\g^2)]/
[\b^2\d(1+\a^2+\g^2)] \\
0&1&1&\g&0& \g(\a^2+\b^2)[(1+\a^2)(1+\g^2+\d^2)+\b^2(\g^2+\d^2)]/
[\b^2(\d^2+(\a^2+\b^2)(\g^2+\d^2))] \\
0&-\g&0&1&\d&1
\end{array} \right],
\eea by appending this new product state.

The symmetric group $S_6$ permutes the 6 product states and induces
a permutation representation on the 720 quintuples made up from
these 6 product states. For instance, if we choose the quintuple
corresponding to column numbers 3,6,2,1,5 (in that order) then the
invariants are
\begin{eqnarray}
J_1^A &=& \frac {(1+\a^2+\g^2)(1+\g^2+\d^2)}
{(1+\g^2)[(1+\a^2)(1+\g^2+\d^2)+\b^2(\g^2+\d^2)]}, \\
J_2^A &=& \frac {1+\g^2}{1+\g^2+\d^2}, \\
J_3^A &=& \frac {(1+\a^2)(1+\g^2+\d^2)+\b^2(\g^2+\d^2)}
{1+\a^2+\g^2}, \\
J_1^B &=& -\frac {\b^2(\g^2+\d^2)(1+\a^2+\g^2)}
{\a^2\g^2[(1+\a^2)(1+\g^2+\d^2)+\b^2(\g^2+\d^2)]}, \\
J_2^B &=& -\frac {\a^2(1+\a^2+\b^2)} {\b^2}, \\
J_3^B &=& \frac{\g^2[(1+\a^2)(1+\g^2+\d^2)+\b^2(\g^2+\d^2)]}
{(\g^2+\d^2)(1+\a^2+\b^2)(1+\a^2+\g^2)}
\end{eqnarray}
and the associated symbol is $ppPNNp$.

A brute force computation shows that there are only 12 different
symbols that belong to these 720 quintuples, namely the UPB symbols
listed in Table 1. This completes the proof.
 \epf

It is easy to see that all 144 symbols arise from some real
quintuples of product states. Hence, apart from PPTES, the NPT
states may also have 5 dimensional kernels with exactly 6 product
states in general position.

By means of Theorem \ref{thm:UPB-type}, we can operationally decide
the UPB-type of 5-dimensional subspaces. This is the key tool in our
proof of Conjecture \ref{conj:HipLMS} in the next section.

\section{\label{sec:product=3x3rank4PPT}
Description of $3\times3$ PPT states of rank 4}

We shall first analyze the 5-dimensional subspaces which arise
as kernels of $3\times3$ PPT states of rank 4. This will help us to
resolve a conjecture proposed in \cite{lms10}.

\subsection{Product states in the kernel of $3\times3$ PPT states
of rank 4}

We need the following lemma which we proved
recently in \cite[Lemma 20]{cd11}.
\bl \label{le:3x3rank4PPTES}
Let $\r$ be a $3\times N$ state such that for some $\ket{a}\in\cH_A$,
$\rank \bra{a}\r\ket{a}=1$. If $\r$ is NPT then it is distillable.
If $\r$ is PPT and $N=3$, then $\r$ is separable.
\el

Let us first handle the separable states of rank 4.
 \bl \label{le:separablekernel}
Let $\r=\sum^3_{i=0} \proj{a_i,b_i}$ be a separable $3\times3$ state
of rank 4. If the four product states $\ket{a_i,b_i}$ are not in
general position, then $\ker\r$ contains a 2-dimensional subspace
$V\ox W$ with $V\subseteq\cH_A$ and $W\subseteq\cH_B$ (which
consists of product states). Otherwise (a) $\ker\r$ contains exactly
6 product states, and (b) these 6 product states are not in general
position.
\el
 \bpf
Assume that the $\ket{a_i,b_i}$ are not in general position. We
consider first the case where two of the $\ket{a_i}$ or two of the
$\ket{b_i}$ are parallel, say $\ket{a_0}\propto\ket{a_1}$. If
$\ket{y}\in\cH_B$ is orthogonal to $\ket{b_2}$ and $\ket{b_3}$, then
$\ket{a_0}^\perp \ox \ket{y}\subseteq\ker\r$. In the remaining case
we may assume that, say $\ket{a_0}$, $\ket{a_1}$ and $\ket{a_2}$ are
linarly independent while $\ket{b_2}$ belongs to the span of
$\ket{b_0}$ and $\ket{b_1}$. If the state $\ket{y}\in\cH_B$ is
orthogonal to $\ket{b_0}$ and $\ket{b_1}$, then $\ket{a_3}^\perp \ox
\ket{y}\subseteq\ker\r$ and so the first assertion is proved.

Next assume that the $\ket{a_i,b_i}$ are in general position. By the
Four Point Lemma, we may assume that these product states are in the
canonical form, i.e., $\ket{a_i,b_i}\propto\ket{ii}$, for $i=0,1,2$
and $\ket{a_4,b_4}=\sum_{i,j=0}^2 \ket{ij}$. The six product states
$\ket{i} \ox (\ket{j}-\ket{k})$ and $(\ket{j}-\ket{k}) \ox \ket{i}$,
with $(i,j,k)$ a cyclic permutation of $(0,1,2)$, belong to
$\ker\r$. To prove (a) one just needs to verify that there are no
additional product states in $\ker\r$. We omit the details of this
verification. For (b), it suffices to observe taht the sum of
$\ket{0}-\ket{1}$, $\ket{1}-\ket{2}$ and $\ket{2}-\ket{0}$ is 0.
This completes the proof.
 \epf

From now on we focus on the PPTES of rank 4. We recall from
\cite{cd11} that any state $\r$ of rank 4 acting on a $3\ox3$
system can be written as
\bea
\label{Rang-4} \r=\sum_{i,j=0}^2 \ket{i}\bra{j}\ox C_i^\dag C_j,
\eea
where the blocks $C_i$ are $4\times3$ matrices.

 \bt \label{thm:3x3rank4PPTES,6productstates}
The kernel of any $3\times3$ PPTES of rank 4 contains exactly
six product states. Moreover, these six states are in general position.
 \et
 \bpf
Let $\r$ be a PPTES of rank 4. Assume that $\ker\r$ contains
infinitely many product states.
By Lemma \ref{le:range=1prodstate} they are all
in general position.
Hence we can apply Proposition \ref{prop:5generalposition}
to any quintuple of product states in $\ker\r$.
As $\ker\r$ contains infinitely many product states, only
the case (a) of that proposition applies. The third assertion of
that case contradicts our hypothesis that $\r$ is PPT. This
contradiction shows that $\ker\r$ contains only finitely many
product states.

We may assume that $\r$ is written as in Eq. (\ref{Rang-4}). We have
$\r=C^\dag C$, where $C=[C_0~C_1~C_2]$. We shall simplify this
expression by using the techniques similar to those in \cite{cd11}.
We can replace $C$ with $UC$ where $U$ is a unitary matrix, without
changing $\r$. The effect of a local transformation $\r\to(I\ox
B)^\dag~\r~(I\ox B)$ is to replace each $C_i$ by $C_iB$. Similarly,
a local transformation $\r\to(A\ox I)^\dag~\r~(A\ox I)$ acts on $\r$
via block-wise linear operations, such as adding a linear
combination of $C_0$ and $C_1$ to $C_2$, etc. We can apply these
kind of transformations repeatedly as many times as needed. Note
also that if the $j$th column of $C_i$ is 0 then $\ket{ij}\in\ker\r$.
By Lemma \ref{le:3x3rank4PPTES}, each block
$C_i$ has rank at least 2.

Since $\ker\r$ has dimension 5, it must contain a product state. We
choose an arbitrary product state in $\ker\r$. By changing the o.n.
bases of $\cH_A$ and $\cH_B$, we may assume that the chosen product
state in $\ker\r$ is $\ket{00}$. Since
$0=\r\ket{00}=\sum_i \ket{i}\ox C_i^\dag C_0\ket{0}$, we must have
$C_i^\dag C_0\ket{0}=0$ for each $i$.
In particular, $C_0^\dag C_0\ket{0}=0$ which implies that
$C_0\ket{0}=0$, i.e., the first column of $C_0$ must be 0.
Hence, the block $C_0$ must have rank $2$, and we may assume that
 \bea
C_0= \left[\begin{array}{ccc}
0 & 1 & 0 \\
0 & 0 & 1 \\
0 & 0 & 0 \\
0 & 0 & 0
\end{array}\right],\quad
C_1=[b_{ij}], \quad C_2=[c_{ij}].
 \eea

Let $\s=\r^\G$ and observe that its first entry is 0. Since
$\s\ge0$, the first row of $\s$ must be 0. We deduce that
$b_{11}=b_{21}=c_{11}=c_{21}=0$. Since $\ker\r$ contains only
finitely many product states, the first columns of $C_1$ and $C_2$
must be linearly independent. By using
an ILO on system A, we may assume that $b_{31}=c_{41}=1$ and
$b_{41}=c_{31}=0$. Thus we have
 \bea
C_1= \left[\begin{array}{ccc}
0 & b_{12} & b_{13} \\
0 & b_{22} & b_{23} \\
1 & b_{32} & b_{33} \\
0 & b_{41} & b_{43}
\end{array}\right],\quad
C_2= \left[\begin{array}{ccc}
0 & c_{12} & c_{13} \\
0 & c_{22} & c_{23} \\
0 & c_{32} & c_{33} \\
1 & c_{42} & c_{43}
\end{array}\right]
 \eea
while $C_0$ did not change.

Let $\cP^4$ be the 4-dimensional complex projective space associated
to $\ker\r$. The Segre variety $\S_{2,2}$ and this $\cP^4$ intersect
properly and we claim that the intersection multiplicity at the
point $P=\ket{00}$ is 1. To prove this claim, we introduce the
homogeneous coordinates $\x_{ij}$ for the projective space $\cP^8$
associated to $\cH$: If $\ket{\psi}=\sum_{i,j=0}^2 \a_{ij}\ket{ij}$
then the homogeneous coordinates of the corresponding point
$\ket{\psi}\in\cP^8$ are $\x_{ij}=\a_{ij}$. The computation will be
carried out in the affine chart defined by $\x_{00}\ne0$ which
contains the point $P=\ket{00}$. We introduce the affine coordinates
$x_{ij}$, $(i,j)\ne(0,0)$, in this affine chart by setting
$x_{ij}=\x_{ij}/\x_{00}$. Thus $P$ is the origin, i.e., all of its
affine coordinates $x_{ij}=0$.

The computation of the intersection multiplicity is carried out in
the local ring, say $R$, at the point $P$. This local ring consists
of all rational functions $f/g$ such that $g$ does not vanish at the
origin, i.e., $f$ and $g$ are polynomials (with complex
coefficients) in the 8 affine coordinates $x_{ij}$ and $g$ has
nonzero constant term. By expanding these rational functions in the
Taylor series at the origin, one can view $R$ as a subring of the
power series ring $\bC[[x_{ij}]]$ in the 8 affine coordinates
$x_{ij}$. We denote by $\gm$ the maximal ideal of $R$ generated by
all $x_{ij}$.

The range of $\r$ is the 4-dimensional subspace spanned by the pure
states $\ket{\psi_i}$, $i=1,\ldots,4$, given by the four columns
of $C^\dag$ (see the proof of Proposition 6 in \cite{cd11}).
In the matrix notation, these pure states are represented
by the following matrices
 \bea
\left[\begin{array}{ccc}
0 & 1 & 0 \\
0 & b_{12}^* & b_{13}^* \\
0 & c_{12}^* & c_{13}^*
\end{array}\right],\quad
\left[\begin{array}{ccc}
0 & 0 & 1 \\
0 & b_{22}^* & b_{23}^* \\
0 & c_{22}^* & c_{23}^*
\end{array}\right],\quad
\left[\begin{array}{ccc}
0 & 0 & 0 \\
1 & b_{32}^* & b_{33}^* \\
0 & c_{32}^* & c_{33}^*
\end{array}\right],\quad
\left[\begin{array}{ccc}
0 & 0 & 0 \\
0 & b_{42}^* & b_{43}^* \\
1 & c_{42}^* & c_{43}^*
\end{array}\right].
 \eea
Since $\ker\r=\cR(\r)^\perp$,
the subspace $\cP^4$ is the zero set of the ideal $I_1$
generated by the four linear polynomials:
 \bea
\label{lin-pol} &&
x_{01}+b_{12}x_{11}+b_{13}x_{12}+c_{12}x_{21}
    +c_{13}x_{22},  \\
&& x_{02}+b_{22}x_{11}+b_{23}x_{12}+c_{22}x_{21}
    +c_{23}x_{22},  \\
&& x_{10}+b_{32}x_{11}+b_{33}x_{12}+c_{32}x_{21}
    +c_{33}x_{22},  \\
&& x_{20}+b_{42}x_{11}+b_{43}x_{12}+c_{42}x_{21}
    +c_{43}x_{22}.
 \eea
The piece of the Segre variety contained in our affine chart
consists of all matrices
 \bea
\left[\begin{array}{ccc}
1 & x_{01} & x_{02} \\
x_{10} & x_{11} & x_{12} \\
x_{20} & x_{21} & x_{22}
\end{array}\right],
 \eea
of rank 1. It is the zero set of the ideal $I_2$ generated by
the four polynomials:
 \bea \label{Segre-pol}
x_{11}-x_{01}x_{10},~x_{12}-x_{02}x_{10},~
x_{21}-x_{01}x_{20},~x_{22}-x_{02}x_{20}.
 \eea 
The quotient space $\gm/\gm^2$ is an 8-dimensional vector space 
with the images of the $x_{ij}$ as its basis. 
It is now easy to see that the images of the generators of $I_1$ 
and $I_2$ also span the space $\gm/\gm^2$.
Hence, by Nakayama's Lemma (see \cite[p. 225]{Cox}) we have
$I_1+I_2=\gm$. 
Consequently, $R/(I_1+I_2)\cong\bC$ and so our claim is proved.

Recall that we chose in the beginning an arbitrary product state
in $\ker\r$ and by changing the coordinates we were able to
assume that this product state is $\ket{00}$. Since the
intersection multiplicity is invariant under these coordinate
changes, this means that we have shown that the intersection
multiplicity is 1 at each intersection point of $\cP^4$ and $\S_{2,2}$.
By B\'{e}zout Theorem the sum of the multiplicities at all
intersection points is 6, and since all of the multiplicities are
equal to 1 we conclude that the intersection consits of exactly 6
points.

By Lemma \ref{le:range=1prodstate}, these six product states are in
general position. This concludes the proof.
 \epf

\subsection{$\G$-invariant PPTES of rank 4}
We shall prove that every SLOCC-equivalence class in $\cE_4$
contains a state which is $\G$-invariant.
  \bt
 \label{thm:invariantPPTES}
Any $3\times3$ PPTES $\r$ of rank 4 is SLOCC-equivalent to one which is
invariant under partial transpose, i.e., for some $(A,B)\in\GL$ and
$\s=A \ox B~\r~A^\dg \ox B^\dg$ we have $\s^\G=\s$.
 \et
 \bpf
By Theorem \ref{thm:3x3rank4PPTES,6productstates}, we may assume
that $\ket{ii}\in\ker\r$ for $i=0,1,2$. Hence we may assume that, in
the formula (\ref{Rang-4}) for $\r$, the column $i+1$ of the block
$C_i$ vanishes for $i=0,1,2$. By multiplying $C=[C_0~C_1~C_2]$ by a
unitary matrix on the left hand side and by performing an ILO with
diagonal matrices we may assume that
 \bea
 C_0= \left[\begin{array}{ccc}
0 & 1 & b \\
0 & 0 & 1 \\
0 & 0 & 0 \\
0 & 0 & 0
\end{array}\right],\quad
 C_1= \left[\begin{array}{ccc}
b_{11} & 0 & b_{13} \\
b_{21} & 0 & b_{23} \\
b_{31}  & 0 & b_{33} \\
b_{41}  & 0 & b_{43}
\end{array}\right],\quad
 C_2= \left[\begin{array}{ccc}
c_{11} & c_{12} & 0 \\
c_{21} & c_{22} & 0 \\
c_{31} & c_{32} & 0 \\
c_{41} & c_{42} & 0
\end{array}\right].
 \eea
Let $\s=\r^\G$ and observe that its first, fifth  and ninth diagonal
entries are 0. Since $\s\ge0$, the rows of $\s$ containing these
entries must be 0. We deduce that $b_{11}$, $b_{21}$, $c_{11}$,
$c_{21}$, $b_{13}$ are 0 and that the second column of $C_2$ is
orthogonal to the first and third columns of $C_1$. Since $\ker\r$
contains only finitely many product states (up to scalar multiple),
the first columns of $C_1$ and $C_2$ must be linearly independent.
Hence, by applying a unitary transformation to the last two rows of
the $C_i$ and rescaling $C_1$ and $C_2$, we may assume that
$b_{31}=0$, $b_{41}=1$ and $c_{31}=1$. By the orthogonality property
mentioned above, we conclude that $c_{42}=0$. Thus we have
 \bea
 C_0= \left[\begin{array}{ccc}
0 & 1 & b \\
0 & 0 & 1 \\
0 & 0 & 0 \\
0 & 0 & 0
\end{array}\right],\quad
 C_1= \left[\begin{array}{ccc}
0 & 0 & 0 \\
0 & 0 & b_{23} \\
0 & 0 & b_{33} \\
1 & 0 & b_{43}
\end{array}\right],\quad
C_2= \left[\begin{array}{ccc}
0 & c_{12} & 0 \\
0 & c_{22} & 0 \\
1 & c_{32} & 0 \\
c_{41} & 0 & 0
\end{array}\right].
 \eea

Since $\r$ is entagled and PPT, its range contains no product
states. Consequently, the entries $c_{12}$, $c_{22}$, $b_{33}$,
$b_{43}$ and $b$ must be nonzero. We choose a phase factor $z_1$
such that $c_{41}z_1>0$. By multiplying the first columns of all
$C_i$ by $z_1$ and mutiplying $C_1$ by $z_1^*$, we may assume that
$d:=c_{41}>0$. By multiplying the last two columns of all $C_i$ with
$1/b_{33}$ and multiplying $C_0$ by $b_{33}$, we may assume that
$b_{33}=1$. We choose the phase factor $z_2$ such that $bz_2>0$. By
multiplying the second columns of all $C_i$ by $z_2^*$ and
mutiplying $C_0$ by $z_2$, we may assume that $b>0$. Since the last
row of $\s$ must vanish, we obtain that $b_{43}=-1/d$,
$c_{12}=-c_{22}/b$ and $c_{32}=-b_{23}^*c_{22}$. If $b_{23}=0$ then
also $c_{32}=0$ and $\ket{\psi_4}-d\ket{\psi_3}$ is a product state
(see Eq. (\ref{Rang-4})). Hence, $c:=b_{23}\ne0$. Next choose a
phase factor $z_2$ such that $cz_2>0$. By multiplying the last two
columns of all $C_i$ by $z_2$ and multiplying $C_1$ by $z_2^*$, we
may assume that $c>0$. Finally, by multiplying the second columns of
all $C_i$ by $a:=1/c_{22}$ we have
 \bea \label{ea:Blokovi}
 C_0= \left[\begin{array}{ccc}
0 & a & b \\
0 & 0 & 1 \\
0 & 0 & 0 \\
0 & 0 & 0
\end{array}\right],\quad
 C_1= \left[\begin{array}{ccc}
0 & 0 & 0 \\
0 & 0 & c \\
0 & 0 & 1 \\
1 & 0 & -1/d
\end{array}\right],\quad
 C_2= \left[\begin{array}{ccc}
0 & -1/b & 0  \\
0 & 1 & 0  \\
1 & -c & 0 \\
d & 0 & 0
\end{array}\right].
\eea
We claim that $a$ has to be real. For that purpose we compute
the principal minor of $\s$ obtained by deleting the first, fifth
and ninth rows and columns. We obtain the expression $c^2(a-a^*)^2$.
Since this minor must be nonnegative, our claim is proved.

It is now easy to verify that $\r^\G=\r$, which completes the proof.
 \epf

From the proof of Theorem \ref{thm:invariantPPTES} it follows that
any PPTES $\r$ is SLOCC-equivalent to one given by Eq. (\ref{Rang-4})
where the blocks $C_i$ are given by Eq. (\ref{ea:Blokovi}).
Moreover the parameters $a,b,c,d$ are nonzero
real numbers with $b,c$ and $d$ positive. The converse also holds
(and it is easy to verify), i.e., if $\r$ is given by this recipe
then it must be a PPTES. The above formulae (\ref{ea:Blokovi})
for the blocks $C_i$ play an essential role in the proof of
the next two theorems.

\bt \label{thm:SameRange} If $\r,\r'\in\cE_4$ have the same range,
then $\r=\r'$. \et

\bpf Clearly, in order to prove the theorem we can simultaneously
transform $\r$ and $\r'$ by the same ILO. Thus we can assume that
$\r$ is given by Eq. (\ref{Rang-4}) and that the blocks $C_i$ are as
in Eq. (\ref{ea:Blokovi}). For convenience, we shall not normalize
neither $\r$ nor $\r'$, and so we have to prove that $\r'$ is a
scalar multiple of $\r$. The range of $\r$ is spanned by the four
pure states $\ket{\ps_i}$ represented by the four matrices
 \bea \label{CistaSt}
\left[\begin{array}{ccc}
0 & a & b \\
0 & 0 & 0 \\
0 & -1/b & 0
\end{array}\right],\quad
\left[\begin{array}{ccc}
0 & 0 & 1 \\
0 & 0 & c \\
0 & 1 & 0
\end{array}\right],\quad
\left[\begin{array}{ccc}
0 & 0 & 0 \\
0 & 0 & 1 \\
1 & -c & 0
\end{array}\right],\quad
\left[\begin{array}{ccc}
0 & 0 & 0 \\
1 & 0 & -1/d \\
d & 0 & 0
\end{array}\right],
 \eea
respectively. The first of these matrices is made up in the obvious
manner from the first rows of the blocks $C_i$, and the other three
matrices are constructed similarly. All this information is
encapsulated in the matrix $C=[C_0~C_1~C_2]$.

We also have $\r'=\sum_{i=0}^3 \proj{\psi'_i}$ where $\ket{\psi'_i}$
are four linearly independent linear combinations of the pure states
$\ket{\psi_i}$. Hence, the matrix $C'$ corresponding to $\r'$ is
given by $C'=SC$, where $S$ is an invertible matrix of order 4. Thus
we have $\r'=C'^\dag HC'$ where $H=S^\dag S$ is a positive definite
matrix of order 4. If $\s'=(\r')^\G$, an easy computation shows that
the first, fifth and ninth diagonal entries of $\s'$ are 0.
Consequently all entries of $\s'$ in the first and fifth row must be
equal to 0. These equations give immediately that $H$ is a scalar
matrix, which completes the proof. \epf

\subsection{Main result}
We can now prove our main result, i.e. Conjecture \ref{conj:HipLMS}.

 \bt
 \label{thm:maintheorem=PPT3x3rank4}
Up to normalization, any state $\r\in\cE_4$ has the form
$A\ox B~\Pi\{\psi\}~ A^\dag\ox B^\dag$ for some $(A,B)\in\GL$ and
some $\{\psi\}\in\cU$.
 \et

\bpf The proof is based on Theorem \ref{thm:UPB-type}. The first
step is to construct the 6 product states in $\ker\r$. Like in
Theorem \ref{thm:SameRange}, we assume that $\r$ is given by Eq.
(\ref{Rang-4}) and that the blocks $C_i$ are as in Eq.
(\ref{ea:Blokovi}). A direct computation shows that the
5-dimensional space $\ker\r$ is spanned by $\ket{00}$, $\ket{11}$,
$\ket{22}$, and any two of the pure product states
 \bea
\label{3MatRang1} \ket{\psi_i}=\left[\begin{array}{ccc}
\left((1+b^2+b^2c^2)z_i-b^2c\right)/abz_i  &
\left((1+b^2+b^2c^2)z_i-b^2c\right)/ab & c-(1+c^2)z_i \\
(cz_i-1-d^2)/d & z_i(cz_i-1-d^2)/d & cz_i-1 \\
1 & z_i & d(cz_i-1)/(cz_i-1-d^2)
\end{array}\right],
 \eea
where $z_i$ $(i=1,2,3)$, are the three roots of
 \bea
\label{kub-jed}
abz(cz-1-d^2)(c-(1+c^2)z)+d(cz-1)(b^2c-(1+b^2+b^2c^2)z)=0.
 \eea
The l.h.s. is a cubic polynomial, say $f(z)$, in the unknown $z$. We
have \bea
&& f(0)=-b^2cd<0,  \\
&& f(\frac{c}{1+c^2})=\frac{c d}{(1+c^2)^2}>0,  \\
&& f(\frac{1+d^2}{c})=
\frac{d^3}{c}\cdot(b^2c^2-(1+b^2+b^2c^2)(1+d^2))<0. \label{nejed}
\eea Since $0<c/(1+c^2)<(1+d^2)/c$, we deduce that Eq.
(\ref{kub-jed}) has three distinct nonzero real roots. One of them
is  $z_1\in(0,c/(1+c^2))$, the second $z_2\in(c/(1+c^2),(1+d^2)/c)$,
and the third is $z_3<0$ if $a>0$ and $z_3>(1+d^2)/c$ if $a<0$. We
also have
 \bea
&& f(1/c)=abd^2/c^2\ne0,\\
&& f(\lambda)=-ab^3c^2\cdot
\frac{(1+b^2)(1+d^2)+b^2c^2d^2}{(1+b^2+b^2c^2)^3}\ne0,
 \eea
where $\lambda=b^2c/(1+b^2+b^2c^2)$. Hence, for $a>0$ we have
 \bea
z_3<0,\quad \lambda<z_1<c/(1+c^2),\quad 1/c<z_2<(1+d^2)/c,
 \eea
and for $a<0$
 \bea
0<z_1<\lambda,\quad c/(1+c^2)<z_2<1/c<(1+d^2)/c<z_3.
 \eea

\begin{figure}
  \includegraphics[width=7cm]{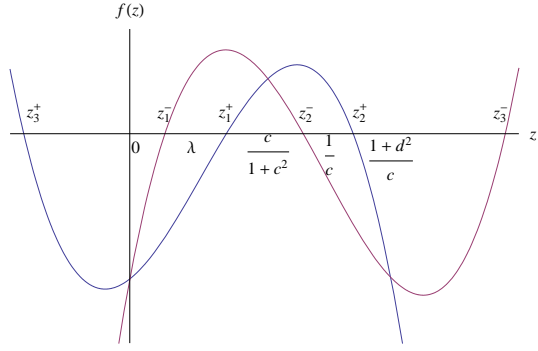}\\
  \caption{\label{fig:Thm21} A generic picture for the function
$f(z)$ for positive and negative $a$, common $b,c,d>0$ and
$\lambda=b^2c/(1+b^2+b^2c^2)$. The left curve represents $f(z)$ for
$a>0$ and the right curve for $a<0$. The two curves meet at three
points with abscisae $z=0,~c/(1+c^2),~(1+d^2)/c$. The three roots of
$f(z)$ are $z_1^-,z_2^-,z_3^-$ for $a<0$, and $z_1^+,z_2^+,z_3^+$
for $a>0$. }
\end{figure}

The second step is to compute the invariants for one of the
quintuples selected from the above 6 product states. For that
purpose we shall use the quintuple
$(\ket{00},\ket{11},\ket{22},\ket{\psi_1},\ket{\psi_2})$. A
computation gives the formulae:
\begin{eqnarray}
&& J_1^A=\frac{1+d^2-cz_2}{1+d^2-cz_1}, \quad
J_2^A=\frac{z_2(z_1-\lambda)}{z_1(z_2-\lambda)}, \quad
J_3^A=\frac{z_1(1+d^2-cz_1)(z_2-\lambda)}
{z_2(1+d^2-cz_2)(z_1-\lambda)}, \\
&& J_1^B=\frac{z_2(1+d^2-cz_2)(cz_1-1)}{z_1(1+d^2-cz_1)(cz_2-1)},
\quad J_2^B = \frac{(1+d^2-cz_1)(cz_2-1)}{(1+d^2-cz_2)(cz_1-1)},
\quad J_3^B = \frac{z_1}{z_2}.
\end{eqnarray}

As the third step, we have to compute the symbol associated to this
quintuple. There are two cases to consider according to whether
$a>0$ or $a<0$. We claim that the symbol is $ppPNNp$ in the former
case and $pNNPpp$ in the latter case. This verification is of
routine nature and Figure 1 may be useful for that purpose. We shall
just verify the claim in the case $a>0$. In that case, it is obvious
that $0<J_1^A<1$, $J_1^B<0$, $J_2^B<0$,  and $0<J_3^B<1$. Since
$\lambda<z_1<z_2$ and the function $t/(t-\lambda)$ is decreasing for
$t>\lambda$, we indeed have $0<J_2^A<1$. Since $J_1^AJ_2^AJ_3^A=1$,
we deduce that $J_3^A>1$. Thus we have shown that the symbol is
$ppPNNp$ if $a>0$.

Finally, since $ppPNNp$ and $pNNPpp$ are UPB symbols, we conclude
that $\ker\r$ is a 5-dimensional subspace of UPB type. Hence, we can
now apply Theorem \ref{thm:SameRange} to complete the proof.
 \epf

It follows from the above proof that every $3\times3$ PPTES
of rank 4 is SLOCC-equivalent to one given by Eqs. (\ref{Rang-4})
and (\ref{ea:Blokovi}) with positive $a,b,c,d$.

The next corollary shows that there is no way to single out one of
the six product states in the kernel of a PPTES of rank 4
in the sense that any quintuple of these states is BP-equivalent
to a qunituple formed from the five states of a UPB.

\bcr \label{cor:5-from-6} Let $\r\in\cE_4$ and let $\ket{\psi_k}$,
$k=0,\ldots,4$, be any five of the six product states in $\ker\r$.
Then there exists $(A,B)\in\GL$ such that the product states $A\ox
B~\ket{\psi_k}$, $k=0,\ldots,4$, form a non-normalized UPB. \ecr
\bpf By Theorem \ref{thm:maintheorem=PPT3x3rank4}, $\ker\r$ is a
5-dimensional subspace of UPB type. By Theorem \ref{thm:UPB-type},
the symbol of the quintuple $(\ket{\psi})=(\ket{\psi_k})_{k=0}^4$ is
a UPB symbol and there exists a permutation $(k_0,\ldots,k_4)$ of
the indexes $(0,\ldots,4)$ such that the symbol of the quintuple
$(\ket{\psi'}):=(\ket{\psi_{k_i}})_{i=0}^4$ is $NPNPpP$. We can now
conclude the proof by using the argument from the last paragraph of
the proof of Theorem \ref{thm:UPB-type}. \epf

We now analyze the stabilizer of $\r\in\cE_4$ in the product
$\PGL=\PGL_3 \times \PGL_3$ of two projective general linear groups.
Thus we have to consider $(A,B)\in\GL$ such that $ A\ox
B~\r~A^\dag\ox B^\dag=c\r$ for some scalar $c>0$.

\bpp \label{Stabilizer} The stabilizer $G_\r$ of any $\r\in\cE_4$ in
$\PGL$ is a finite group isomorphic to a subgroup of the symmetric
group $S_6$. In the generic case the stabilizer is trivial. \epp
\bpf Let us denote by $P_i$, $i=1,\ldots,6$, the six points in the
projective space $\cP^4$ associated with $\ker\r$. Assume that
$(A,B)\in\GL$ maps $\r$ to $c\r$ for some $c>0$. Then $A\ox B$ must
leave invariant $\cR(\r)$ and $(A^{-1}\ox B^{-1})^\dag$ must leave
invariant $\ker\r$ and permute the 6 product states, i.e., the
points $P_i$. The map which assigns to $(A,B)$ this permutation, say
$\pi_{A,B}\in S_6$, is a group homomorphism. If $\pi_{A,B}$ is the
identity permutation, then the Four Point Lemma implies that
$(A^{-1}\ox B^{-1})^\dag$ is a scalar operator. We conclude that the
mapping sending $(A,B)\in\G_\r$ to the permutation $\pi_{A,B}\in
S_6$ is one-to-one. This proves the first assertion.

The proof of the second assertion is based on matrices $U$ and $V$
from the proof of Theorem \ref{thm:UPB-type} and the invariants
$(J_i^A,J_i^B)$, $i=1,2,3$, defined in Section \ref{sec:ProdV}. Let
us write the product states $\ket{\psi_i}\in\ker\r$,
$(i=1,\ldots,6)$ as $\ket{\psi_i}=\ket{\phi_i}\ox\ket{\chi_i}$. The
first five $\ket{\phi_i}$ and $\ket{\chi_i}$ are determined directly
by $U$ and $V$, respectively. One has to compute the sixth product
state in $\ker\r$. We have done this in the proof of the theorem
mentioned above. Finally, we have verified that the 720 (ordered)
quintuples that one can construct from the 6 points $P_i$ have
generically different values for the invariants $(J_i^A,J_i^B)$.
This completes the proof. \epf

For special states $\r\in\cE_4$, the stabilizer may be nontrivial.
For instance, the stabilizer for the {\bf Pyramid} example is the
alternating group $A_5$ of order 60 which permutes transitively
the six product states in the kernel.
The original definition of this example \cite{bds96} exhibits only
the dihedral group of order 10 as symmetries of a regular
pentagonal pyramid. A more symmetric realization is given in the
recent paper \cite{slm11} where the full group of symmetries, $A_5$,
can be realized by local unitary operations.

For the {\bf Tiles} example the stabilizer is a group of order 12
isomorphic to the alternating group $A_4$. It also permutes
transitively all six product states in the kernel of $\r$. Since
this stabilizer is finite, it may be conjugated into the maximal
compact subgroup of $\PGL$ which is just the image of the local
unitary group. In this way we can obtain a more symmetric
realization of {\bf Tiles}. To be concrete, let us consider the
following two $3\times6$ matrices
 \bea
\tilde{U}=\left[\begin{array}{cccccc}
a&0&1&a&0&-1\\1&a&0&-1&a&0\\0&1&a&0&-1&a
\end{array}\right],\quad
\tilde{V}=\left[\begin{array}{cccccc}
-1&0&a&1&0&a\\0&a&-1&0&a&1\\a&-1&0&a&1&0
\end{array}\right],
 \eea
where $a=\root 3 \of 3$. Define the pure product state
$\ket{\psi'_{k-1}}$, $(k=1,\ldots,6)$, as the tensor product of the
$k$th columns of $\tilde{U}$ and $\tilde{V}$. These states are
linearly dependent since the first three and the last three have the
same sum. A computation shows that the {\bf Tiles} quintuple
$\ket{\psi_k}$, $k=0,\ldots,4$, and the quintuple
$(\ket{\psi'_0},\ket{\psi'_1},\ket{\psi'_4},
\ket{\psi'_3},\ket{\psi'_2})$ have the same invariants. Hence, they
are BP-equivalent. The advantage of this new realization of {\bf
Tiles} is that its symmetry group (i.e., the stablizer) is now
evident. The symmetry operations are given by local unitary
operations. For instance, the multiplication of $\tilde{U}$ and
$\tilde{V}$ by $\diag(-1,1,1)$ acts on the 6 product states as the
permutation (03)(25). The multiplication of $\tilde{U}$ by the
cyclic matrix $Z$ with first row $[0~0~1]$ and the simultaneous
multiplication of $\tilde{V}$ by $Z^T$ act as the permutation
(012)(345). One can easily construct the unique PPTES $\r$ of rank 4
whose kernel is the subspace spanned by the $\ket{\psi'_k}$. Up to
normalization, $\r$ is given by Eq. (\ref{Rang-4}) with
 \bea
C_0=\left[\begin{array}{ccc}
a&0&0\\0&0&1\\0&0&0\\0&a^2&0\end{array}\right],\quad
C_1=\left[\begin{array}{ccc}
0&0&a\\a^2&0&0\\0&1&0\\0&0&0\end{array}\right],\quad
C_2=\left[\begin{array}{ccc}
0&a&0\\0&0&0\\0&0&a^2\\1&0&0\end{array}\right].
 \eea It is now easy to
verify that the above two local unitary transformations commute with
$\r$.

\section{ \label{sec:PhysicalApplication}
Physical applications }

In this section we demonstrate a few applications of our results in
previous sections.

First, apart from existing numerical tests such as the
semidefinite programming \cite{dps04}, Theorem
\ref{thm:maintheorem=PPT3x3rank4} analytically provides the first
bipartite system with specified dimensions and rank, in which all
PPTES $\r$ can be systematically built and characterized. The
procedure is as follows. Given a 4-dimensional subspace, one can
readily obtain its 5-dimensional orthogonal complement.
Numerically it is possible to find the product states of this subspace. By Theorems
\ref{thm:UPB-type} and \ref{thm:maintheorem=PPT3x3rank4}, it
suffices to consider the case in which this subspace contains
exactly 6 product states $\ket{\phi_i,\c_i}$, which are
required to be in general position. Next by using Eq.
(\ref{J1-J2}) and condition (b) of Theorem \ref{thm:UPB-type}, we
can compute the invariants $(J_1^A,J_2^A,J_3^A,J_1^B,J_2^B,J_3^B)$
for any quintuple selected from these 6 product states.
It is necessary that all these invariants be real and that their
symbols be of UPB type. Then we compare them with
Eq. (\ref{UPB-inv}) and derive the parameters
$\g_{A,B},\t_{A,B} \in (0,\pi/2)$ of a UPB quintuple $\ket{a_i,b_i}$,
see Lemma \ref{le:domen}. Next, we find the ILOs $A,B$ such that
$A\ox B\ket{a_i,b_i} \propto \ket{\phi_i,\c_i} $.
Finally, we build the PPTES $A \ox B
(I-\sum^4_{i=0}\proj{a_i,b_i})A^\dg \ox B^\dg$, which is unique
by Theorem \ref{thm:SameRange}. This is also succintly stated in
Theorem \ref{thm:maintheorem=PPT3x3rank4}.

On the other hand it is known that the PPT condition is necessary
and sufficient for detecting any $2\times2$ and $2\times3$ separable
states. Likewise, Theorems \ref{thm:UPB-type} and
\ref{thm:maintheorem=PPT3x3rank4} essentially outperform all
existing criteria, such as the range criterion \cite{horodecki97} 
and the covariance matrix criterion \cite{ghg07}, 
which can only detect some special $3\times3$ PPTES of rank 4. 
It also follows easily from Theorem \ref{thm:invariantPPTES} that 
 \bcr
In a two-qutrit system, the partial transpose of a PPTES of rank 4
has also rank 4.
 \ecr

Second, we claim that in a two-qutrit system all PPTES of rank 4
are extreme points of the set $\cS_{\ppt}$ of PPT states. 
(Here our states are assumed to be normalized.)
According to the definition in Sec. I, a PPTES $\r$ is not extreme
if and only if it is the midpoint of the segment joining two
different PPT states.
Let us prove our claim. Suppose that $2\r=\r_1 + \r_2$ where
$\r_1$ and $\r_2$ are two distinct PPT states. Since there is no
product state in the range of $\r$, the same is true for $\r_1$
and $\r_2$. Thus they are PPTES, and $\rank\r_1=\rank\r_2=4$ since
there is no PPTES of rank 2 or 3 \cite{hlv00}. Hence, these three
states  must have the same range, and so $\r_1=\r_2=\r$ by
Theorem \ref{thm:SameRange}. Thus we have a contradiction.
Hence $\r$ is always both extreme and edge PPTES \cite{hkp03}.
The latter statement readily follows from the definition of
the edge PPTES $\r$, which contains no product state
$\ket{a,b}\in\cR(\r)$ such that $\ket{a^*,b}\in\cR(\r^\G)$
\cite{lkc00}.

Third, we can systematically build more PPTES and detect them in
experiment. By following the technique in \cite{ghb02,lkc00}, the
entanglement witness of the PPTES $\s$ of rank 4 has the form
 \bea
 \label{ea:entanglementwitness}
 W_{\s} &=& P - \e I ,
 \nonumber\\
\e &=& \inf_{\ket{e,f}} \bra{e,f} P \ket{e,f},
 \eea
where $P$ is the projector onto the kernel of $\s$. The numerical
estimation of some states $\s$ is also available, as well as their
experimental realization in \cite{ghb02}. In this sense, we can
detect any two-qutrit PPTES of rank 4 effectively.

On the other hand it has been proved that any PPTES has the form
$\r={p}\r_s+(1-p)\s$ where $\r_s$ is a separable state and $\s$ is
an edge PPTES \cite{lkc00}. In the second item, we have shown that
any PPTES of rank 4 is an extreme and edge PPTES \cite{hkp03}. So
the PPTES $\r$ can be characterized when $\s$ has rank 4 and the
perturbation $p$ is small, by using the entanglement witness
$W_{\s}$ in Eq. (\ref{ea:entanglementwitness}). Building the
entanglement witness for arbitrary $\r$ requires the
characterization of separable $\r_s$, which is still an open
problem. Nevertheless, if a PPTES is the convex sum of a few PPT
states of rank at most 4, it is possibly characterized
through the results in this paper.

Fourth, we claim that no PPTES $\r\in\cE_4$ is symmetric, that
is, $\cR(\r)$ is not contained in the space spanned by the
$\ket{ii}$ and the $\ket{ij}+\ket{ji}$ with $i>j$.
To prove this, we use the expression in Eq. (\ref{ea:Blokovi}). 
Assume there is an ILO $B=[b_{ij}]$ such that 
$I \ox B ~\r~ I \ox B^\dg$ is symmetric. 
One can readily show that $b_{11}=b_{21}=b_{20}=b_{22}=0$, which
contradicts the assumption $\det B \ne 0$. Hence the simplest
symmetric PPTES must have rank at least 5. Such states indeed exist,
e.g., the state $\r_{BE4}$ in \cite{tg09}.

Finally, up to ILO, the range of any $3\times3$ PPTES of rank 4 is
orthogonal to a 5-dimensional subspace spanned by a UPB. Hence the
product vectors of a UPB cannot be distinguished by LOCC
\cite{bdf99}. In other words, the LOCC-indistinguishable nonlocality
of the complementary subspace is a deterministic feature of any
PPTES of rank 4.

\section{Conclusions}

In this paper we have shown that any two-qutrit PPT entangled state
of rank 4 is the normalization of
 \bea (A\ox B)(I-\sum^4_{i=0}\proj{a_i,b_i})(A\ox B)^\dag,
 \eea where $A,B$ are
invertible operators and the five product states $\{\ket{a_i,b_i}\}$
form an (orthogonal) UPB. Moreover, this is the only PPT entangled
state among the states having the same range. The 5-dimensional
subspace spanned by a UPB contains exactly 6 product states (up to
scalar multiple). We have shown that any 5 of them can be converted
to a UPB quintuple by a biprojective transformation. The result has
been demonstrated on two well-known examples of UPB, the {\bf
Pyramid} and {\bf Tiles} UPB \cite{bdm99}. Therefore we have
systematically characterized all PPTES in this system.

Furthermore we have characterized the separable two-qutrit states
of rank 4 whose kernel contains either infinitely many product
states or exactly 6 product states but not in general position.
The next goal in the future is to extend our results to higher
dimensional PPT states, entangled or separable.

On the other hand, we have proposed a method of determining the
BP-equivalence between two quintuples of product states of two
qutrits in general position. Apart from the derivation of the main
result, the method has also been applied to classify the
5-dimensional subspaces via their intersection with the set of
product states. In particular all 11 partitions of the integer 6
occur as the intersection patterns of $\cP^4\cap\Sigma_{2,2}$.
These results are useful to characterize the distillability of
NPT states of rank 4, which is another interesting open problem
proposed in \cite{cd11}.

\acknowledgments

We thank an anonymous referee for his suggestions, and
L. Skowronek \cite{skow} for pointing out an
inaccuracy in the original version of the paper.
The CQT is funded by the Singapore MoE and the
NRF as part of the Research Centres of Excellence programme. The
second author was supported in part by an NSERC Discovery Grant.

\end{document}